\documentclass{aa}
\usepackage{booktabs}
\usepackage{mathtools}
\usepackage{natbib}
\usepackage{graphicx}
\usepackage{pgf}
\usepackage{txfonts}
\usepackage{enumitem}

\usepackage{gensymb}
\usepackage{textcomp}
\usepackage{amsmath}
\usepackage{tabularx}

\usepackage{threeparttable}
\usepackage{mathtools}

\usepackage{aas_macros}

\usepackage{soul}
\definecolor{lightgreen}{HTML}{B7F774}
\definecolor{lightred}{HTML}{FF6666}
\definecolor{lightorange}{HTML}{FE9A2E}
\sethlcolor{lightgreen}

\usepackage[normalem]{ulem}

\newcommand{\unit}[1]{\ensuremath{\, \mathrm{#1}}}

\newcommand{\slfrac}[2]{\left.#1\middle/#2\right.}

\begin{document}

  \title{Kinematics of the jet in M87 on scales of \\100 -- 1000 Schwarzschild
  radii}

\author{F. Mertens\inst{1, 2}
        \and
        A. P. Lobanov\inst{1,3}
        \and
        R. C. Walker\inst{4}
        \and
        P. E. Hardee\inst{5}
         }

\institute{Max-Planck-Institut f\"ur Radioastronomie,
          Auf dem Hugel 69, 53121 Bonn, Germany
          \and
          Kapteyn Astronomical Institute, 
          PO Box 800, 9700 AV Groningen, The Netherlands
          \and
          Institut f\"ur Experimentalphysik, Universit\"at Hamburg, 
          Luruper Chaussee 149, 22761 Hamburg, Germany
          \and
          National Radio Astronomy Observatory, Socorro, NM 87801, USA
          \and
          Department of Physics \& Astronomy, The University of Alabama,
          Tuscaloosa, AL 35487, USA
          }

\date{}

\abstract
{Very long baseline interferometry (VLBI) imaging of radio emission
  from extragalactic jets provides a unique probe of physical
  mechanisms governing the launching, acceleration, and collimation of
  relativistic outflows.}
{VLBI imaging of the jet in the nearby active galaxy M\,87 enables
  morphological and kinematic studies to be done on linear scales down
  to $\sim 100$ Schwarzschild radii ($R_\mathrm{s}$).}
{The two-dimensional structure and kinematics of the jet in M\,87
  (NGC\,4486) have been studied by applying the Wavelet-based Image
  Segmentation and Evaluation (WISE) method to 11 images obtained from
  multi-epoch Very Long Baseline Array (VLBA) observations made in 
January-August 2007 at 43 GHz ($\lambda = 7$ mm).}
{The WISE analysis recovers a detailed two-dimensional velocity field
  in the jet in M\,87 at sub-parsec scales. The observed evolution of
  the flow velocity with distance from the jet base can be explained
  in the framework of MHD jet acceleration and Poynting flux
  conversion. A linear acceleration regime is observed up to $z_{obs}
  \sim 2$\,mas. The acceleration is reduced at larger scales, which is
  consistent with saturation of Poynting flux conversion.  Stacked
  cross correlation analysis of the images reveals a pronounced
  stratification of the flow. The flow consists of a slow, mildly
  relativistic layer (moving at $\beta \sim 0.5\,c$), associated
  either with instability pattern speed or an outer wind, and a
  fast, accelerating stream line (with $\beta \sim 0.92$, corresponding to
  a bulk Lorentz factor $\gamma
  \sim 2.5$). A systematic difference of the apparent
  speeds in the
  northern and southern limbs of the jet is detected, providing
  evidence for jet rotation. The angular velocity of the magnetic
  field line associated with this rotation suggests that the jet in
  M87 is launched in the inner part of the disk, at a distance $r_0
  \sim 5\, R_\mathrm{s}$ from the central engine.}
{The combined results of the analysis imply that MHD acceleration and
  conversion of Poynting flux to kinetic energy play the dominant
  roles in collimation and acceleration of the flow in M\,87. }


\maketitle

\section{Introduction}

M\,87 (Virgo~A, NGC\,4486, 3C\,274) is a giant elliptical galaxy. Its proximity
($D=16.7$ Mpc; \citeauthor{mei_acs_2007} \citeyear{mei_acs_2007}) combined with
the large mass of its central black hole~\citep[$M_\mathrm{BH} \simeq 6.1 \times
10^9 M_{\odot}$;][] {gebhardt_black_2011}\footnote{The black hole mass
estimates
reported   in the literature for M\,87 are typically in the range of $\sim$3--7
billion  solar masses \citep[cf.][and references   therein]{walsh_m87_2013}.
Throughout this paper, we adopt the   value of $M_{BH} = 6.1\times 10^9
M_{\odot}$.} make M\,87 one of the primary sources to probe jet formation and
acceleration down to smallest linear scales (with $1\ \text{mas} \approx 0.08\
\text{pc} \sim 140\ \text{Schwarzschild radii}~ (R_\mathrm{s})$).


The kiloparsec-scale structure of the jet in M\,87 has been extensively
studied in the radio regime using the Very Large Array (VLA)
\citep{owen_highresolution_1989}, the optical regime using the Hubble
Space Telescope (HST) \citep{biretta_detection_1995}, and the X-ray
regime using the Chandra space telescope
\citep{marshall_highresolution_2002}. These observations have revealed
an edge brightened conical jet with an apparent opening angle
$\Theta_{\mathrm{obs}} \sim 3.2 \degree$ (in this paper, ``opening
angle''
refers
to half the full opening angle). The jet manifests several
bright knots, with a particularly strong feature HST-1 located at a
projected distance of $z_{\mathrm{obs}} \sim 0.84''$. The jet expends
uniformly up to knot A at a distance $z_{\mathrm{obs}} \sim
12''$. Analysis of multi-epoch observations has yielded detections of
superluminal speeds of up to $\beta_ {\mathrm{app}} \sim 6\,c$ in the
optical \citep{biretta_hubble_1999} and $\beta_{\mathrm{app}} \sim
4\,c$ in the radio
\citep{biretta_detection_1995,cheung_superluminal_2007,giroletti_kinematic_2012}
at the location of HST-1, with evidence of deceleration downstream
from this point. Several components with subluminal speed
($\beta_{\mathrm{app}} \sim 0.5\,c$) were also detected all along the
jet \citep{biretta_hubble_1999,meyer_optical_2013}, concurrently with
the detections of superluminal features made at similar distances.

\begin{figure*}
    \centering
    \includegraphics{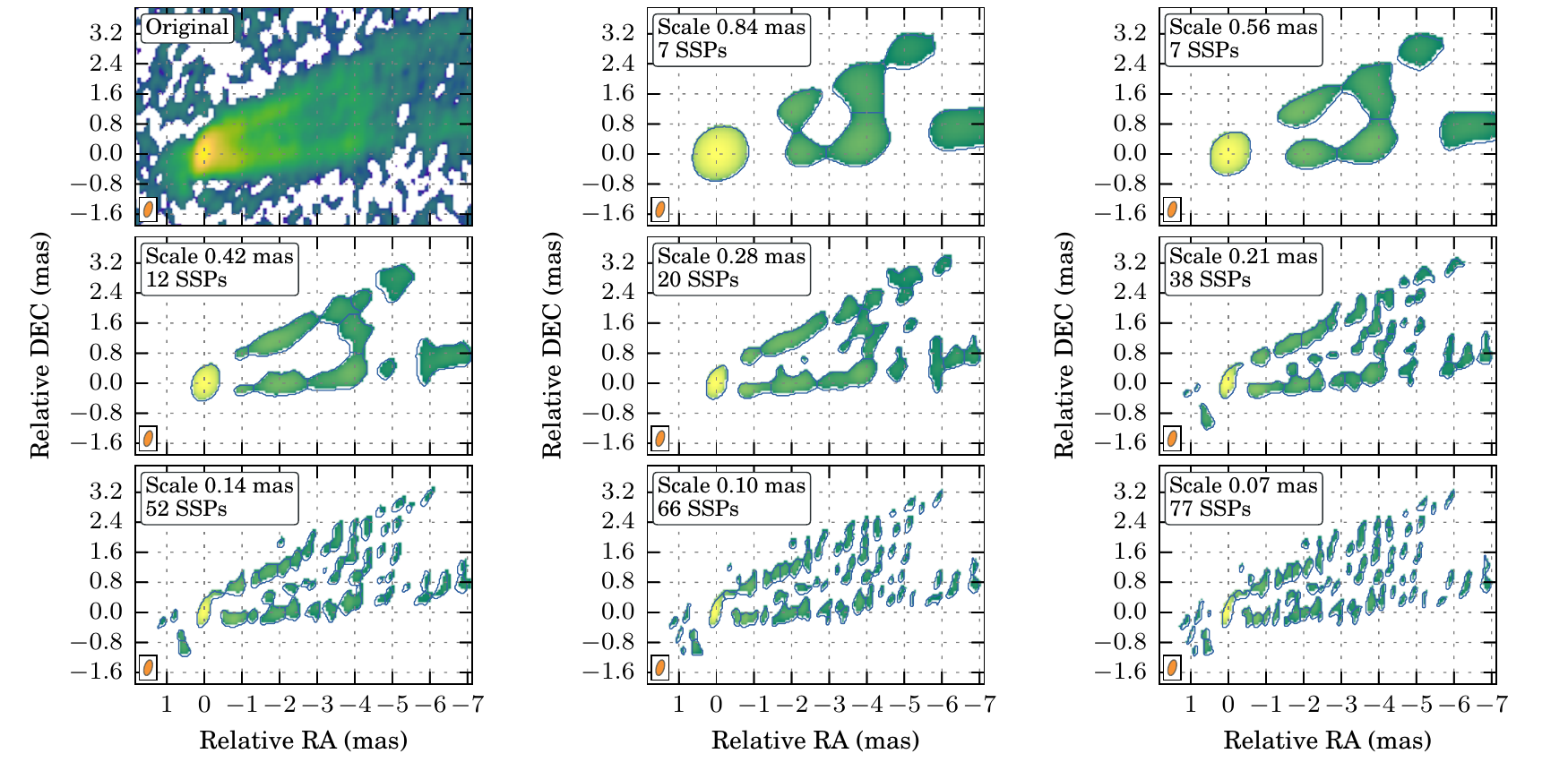}
    \caption[WISE decomposition of the VLBA image of the jet in M\,87 at the
    epoch 2007/02/17.]
    {43\,GHz VLBA image of the jet in M\,87 at the epoch 2007/02/17 (top
    left panel) and its WISE decomposition on SWD scales 1-4 
    (0.07, 0.14, 0.28, 0.56\,mas) and IWD scales (0.105, 0.21, 0.42, 0.84\,mas)
    .}
    \label{fig:m87_wise_decomposition}
\end{figure*}

Both the radio and optical maps exhibit bright filaments, twisted into an
apparent double-helical pattern. This has been successfully modeled as a
consequence of Kelvin Helmholtz (K-H) instability developing in the jet
\citep{lobanov_internal_2003,hardee_using_2011}, with the resulting expected
pattern speed of $\approx 0.5\,c$ that can explain the subluminal features found
in the jet. In the optical regime, jet structure is dominated by knots, while
the filaments are more pronounced in the radio images.  This difference has been
interpreted in the framework of a stratified jet by \cite{perlman_optical_1999},
in which the high energy plasma particles emitting optical synchrotron radiation
originate from regions close to the spine of the flow, while the lower energy
radio emitting particles are concentrated in the outer layers of the flow. 




The physical nature of the inner jet in M\,87 has been the subject
of a number of studies \citep[cf.,][for a
review]{biretta_review_1995}. Recently, multi-frequency
phase-referencing VLBI observations \citep{hada_origin_2011} were used
to locate the central engine at a projected distance of only about
$40\ R_\mathrm{s}$ upstream from the base (or the ``core'') of the radio jet
observed at 43 GHz. Jet expansion can be described with a
parabolic profile~\citep{asada_structure_2012}, indicating that the
jet is collimated by magnetohydrodynamic (MHD) processes
\citep{meier_magnetohydrodynamic_2001}. The transition to a conical
jet was found to occur at a projected distance of $\sim 350$ mas from
the core.


Proper motions detected in the jet in M\,87 at parsec scales have been
the subject of an intensive debate
\citep[cf.,][]{biretta_review_1995,kovalev_inner_2007,ly_high_2007,walker_movie_2009}. The
jet has been observed at 2 cm wavelengths as part of the MOJAVE
project, with a reported slow subluminal speed ($\beta_{\mathrm{app}}
\sim 0.01\,c$) \citep{kovalev_inner_2007} which seems to be in
contradiction with proper motions of up to $\beta_{\mathrm{app}} \sim
4-6\,c$ observed at HST-1 \citep{biretta_hubble_1999}.  Recently, VLBI
observations at 1.6 GHz \citep{asada_discovery_2014} have been used to
claim a detection of gradual acceleration of the jet between 100 mas
and 900 mas, suggesting a link between the subluminal acceleration
found in the first 20 mas of the jet, and the relativistic speed at
HST-1. A major issue persists in this picture: assuming
that the jet is intrinsically bidirectional and symmetric, detection
of a subluminal speed would imply a jet to counter-jet intensity
ratio near unity, while the jet in M\,87 is essentially one sided. The
presence of a counter-jet is suggested from 15\,GHz, 43\,GHz, and
86\,GHz VLBI observations
\citep{kovalev_inner_2007,ly_high_2007,hada_cjet_2016}. However, the
counter-jet emission is only about 2 mas in extent, and jet to
counter-jet ratios of $\sim5$--20 are reported. These findings support
an alternative scenario in which the ultra-slow speed of $\approx
0.01\,c$ is either a pattern speed or a slow wind from an outer sheath
\citep{kovalev_inner_2007}.  This conclusion is further strengthened
by the observed limb-brightened morphology of the jet. Sparsity of the
MOJAVE time sampling is a likely reason for the non-detection of
superluminal speeds in M\,87. The MOJAVE observations of M\,87 are
sparse, with an average time interval of about 200 days between
successive epochs, resulting in a maximum reliably detectable speed of
about $1\,c$ (taken as a displacement of about the beam size between
consecutive epochs).  While \cite{kovalev_inner_2007} could not
identify any fast component, \cite{ly_high_2007} reported possible
features with speeds between $0.25\,c$ and $0.4\,c$ at a distance of 3
mas from the core, with the detection obtained from only 2 epochs.

\subsection{The M\,87 VLBA movie project}

In an attempt to finally assess the true bulk speed of the jet at sub-parsec
scales, systematic VLBA\footnote{Very Long Baseline
Array of the National Radio Astronomy Observatory, USA} observations were performed
by~\cite{walker_vlba_2008} at 43\,GHz ($\lambda =7$\,mm) with a high cadence in
time. The first set of pilot observations indicated that a fast moving component
could indeed be traced. This led to a program that produced a batch of eleven
observations between 27 January 2007 and 26 August 2007, with an average time
interval of 21 days, which allowed detection of apparent speeds as high as
$3$--$4\,c$. The result is a set of excellent-quality, high-resolution VLBI maps
with nearly homogeneous image rms noise. A more complete presentation of
the
2007 data used here along with more densely sampled, but lower quality, data
from 2008 plus roughly annual observations between 1990 and 2016 will be made
separately~\citep[][{\em in prep.}]{walker_m87_2016}

In the images of M\,87 obtained from the VLBA movie project, nearly the entire
jet is well resolved, with up to 3 -- 4 beamwidths across the flow. The 
limb-brightened morphology is striking, especially if one looks at the stacked image
that comprises the eleven observations into a single, high-sensitivity map.
Determination of the velocity of the flow is however difficult from these data.
The standard procedure for making proper motion measurements relies on modeling
the source with a set of Gaussian components and cross-identifying those
components between epochs. In this case, the complexity of the transversely
resolved structure of the jet precludes application of this approach. Visual
inspection was done in an attempt to identify apparently related features and a
speed of about $2\,c$ was consistently found~\citep{walker_vlba_2008}. This
method obviously lacks objectivity and robustness, and it is difficult to use it
for obtaining a complete velocity field map of the jet.


A robust analysis of the M\,87 VLBA movie dataset is required in order
to make a quantitative assessment of the jet morphology and
kinematics. Such an assessment can be made using the wavelet-based
image segmentation and evaluation (WISE) method
\citep{mertens_waveletbased_2015} and the stacked cross-correlation
algorithm \citep{mertens_detection_2016} developed for determining
two-dimensional velocity fields.  In the following a complete WISE
analysis of the internal structure and dynamics of the jet in M\,87 at
sub-parsec scale is presented. In section~\ref{sc:m87:data_analysis},
the WISE method is used to determine the two-dimensional velocity
field, and the collimation profile of the jet from the 43\,GHz VLBA
observations of~\cite{walker_vlba_2008}. The complementary WISE
analysis of the 15\,GHz VLBA observations from the MOJAVE project is
presented in section~\ref{sc:m87:wise_velocity_2cm}. The physical
properties of the jet are discussed in section~\ref{sc:discussion}. In
section~\ref{sc:jet_stratification_model}, we present a physical model
for the flow stratification observed in M\,87. In
section~\ref{sc:m87:viewing_angle}, the viewing angle is derived from
the velocity observed on the counter-jet side. In
section~\ref{sc:m87:jet_rotation}, evidence for jet rotation is
discussed, as inferred from the differential velocity between the
northern and southern limbs of the jet. Based on these findings, the
acceleration and collimation of the jet are discussed in
Sect.~\ref{sc:m87:jet_collimation_acc} in the framework of MHD jet
acceleration. The physical nature of the jet spine is discussed in
Sect.~\ref{sc:nature_spine}, and a plausible jet launching mechanism
is proposed in Sect.~\ref{sc:launching_mechanism}.


Throughout the paper, a cylindrical coordinate system ($z$, $r$, $\phi$) is
adopted for the jet description. In this coordinate system, the coordinates $z$
and $r$ correspond to axial and radial directions in the jet. We set the jet
axis ($z$) at a position angle of $20\degree$ with respect to the RA axis, and
we denote the projected axis as $z_{\mathrm{obs}}$.

\section{WISE analysis of 7 mm VLBA observations}
\label{sc:m87:data_analysis}

We applied WISE analysis \citep{mertens_waveletbased_2015} to 11
images of the jet in M\,87 obtained from 43\,GHz VLBA observations
made between 27 January 2007 and 26 August 2007, with an average time
interval of 21 days between the successive epochs and a common
restoring beam of $0.43 \times 0.21$ mas in PA $-16\degree$. We refer to
\cite{walker_vlba_2008,walker_m87_2016} for a complete description of the data
reduction. At the cadence of the 43\,GHz VLBA measurements, a
displacement comparable to one fifth of the beam corresponds to a
minimum detectable proper motion of 0.74 mas/yr, or $0.19\,c$. The
WISE algorithm has been shown to provide robust detections of up to
$\approx 4$\,FWHM between pairs of successive epochs
\citep{mertens_waveletbased_2015}, which translates into a maximum
apparent speed of $\sim 5\,c$ that can be detected from the VLBA
images of M\,87.  Hence, the WISE analysis should be able to detect
essentially the full range of speeds reported in the compact jet of
M\,87.

\subsection{Velocity field analysis}
\label{sc:m87:wise_velocity_7mm}




Analysis of the M\,87 images is performed as described in
~\cite{mertens_waveletbased_2015}. Each map is decomposed and segmented using
the segmented wavelet decomposition (SWD) method, which provides a description
of the two-dimensional jet structure based on a set of significant structural
patterns (SSP). These features are then cross-identified in pairs of adjacent
epochs using the multiscale cross-correlation (MCC) method. This procedure
yields spatial displacements and proper motions for each of the SSPs identified.

\begin{figure*}
    \centering
    \includegraphics{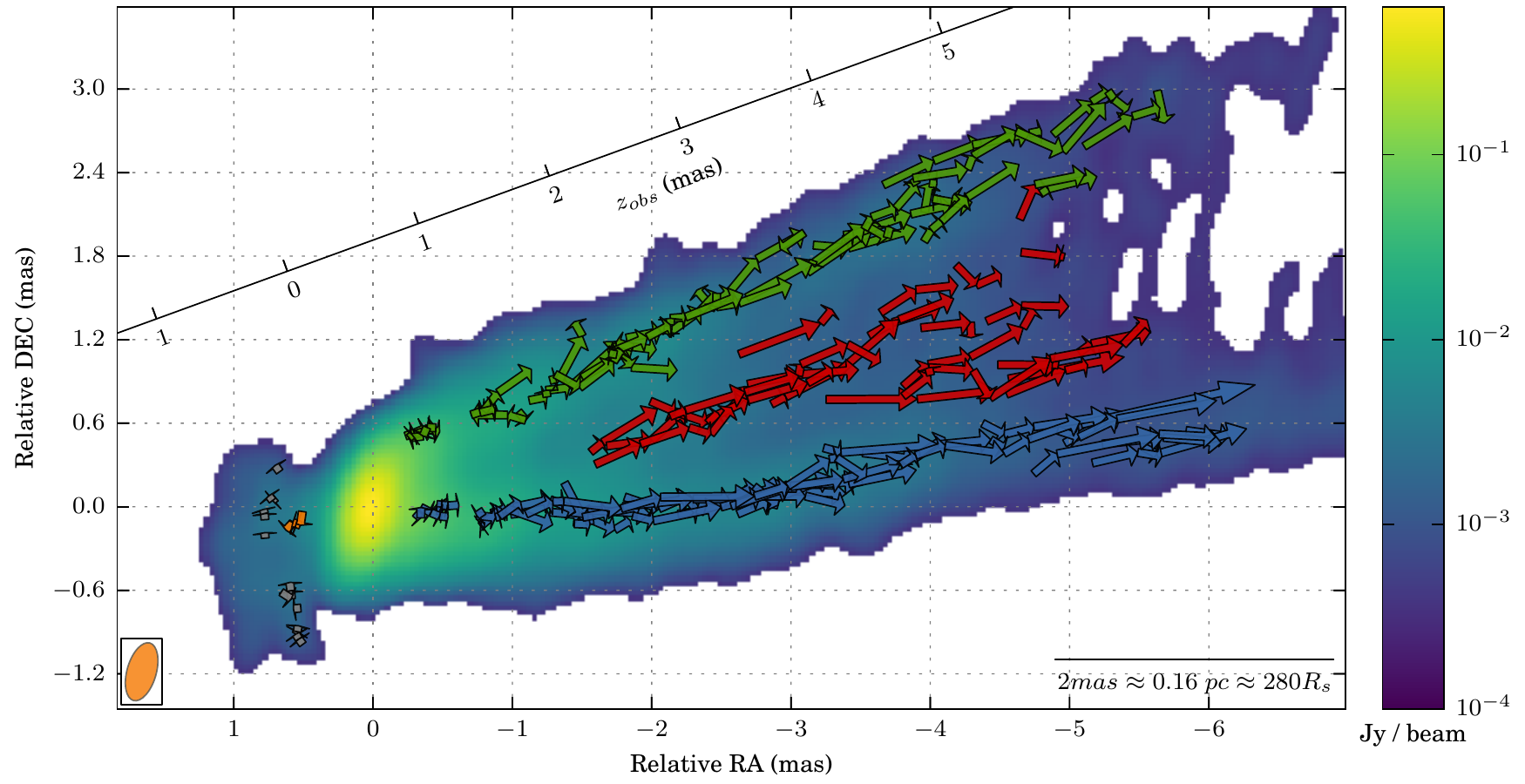}
    \caption[Observed displacements of the significant structural patterns detected in the jet of M\,87.]
    {Observed displacements of the significant structural patterns
      detected in the jet of M\,87. Three main regions can be
      identified in the jet, as revealed by consistency between the
      individual displacements in the southern (blue) and northern (green)
      limbs and in the central ``stream'' (red) of the flow.  The
      components found in the inner part of the counter-jet (orange),
      may reflect motions in a slow, weakly deboosted wind on the
      counter-jet side. In the outer part of the counter-jet (gray), a slower
      transversal motion is evident, with a tendency toward the northward
      direction. The displacements are overplotted on a stacked-image of the
      jet clipped at a 5 $\sigma$ noise rms level (0.5 mJy / beam).}
    \label{fig:m87_velocity_field}
\end{figure*}

To ensure robust detectability of displacements larger than one beamwidth
(corresponding to apparent speeds of $\gtrsim 2\,c$), we   applied the SWD on
four spatial scales (ranged as powers of 2) of   0.07, 0.14, 0.28, and 0.56 mas
and amended them with the   intermediate wavelet decomposition (IWD) performed
on scales of   0.105, 0.21, 0.42, and 0.84 mas.  The IWD implementation,
described in Appendix~\ref{sect:iwd}, is made to improve the efficiency of the
MCC algorithm for recovering structural changes in optically thin, stratified
flows. With these settings, maximum displacements of about 0.84 mas
(corresponding to proper motion of $\sim 18$\,mas/yr and $\beta_{\mathrm{app}}
\sim 4.5\,c$) can be robustly detected in each successive pair of images even
for overlapping optically thin regions with different apparent speeds. The noise
was estimated by computing $\sigma_j$ at each wavelet scale, as described
in~\cite{mertens_waveletbased_2015} and a $3\ \sigma_j$ thresholding was
subsequently applied at each scale. The typical SNR found in different parts
of the jet and for different Scales is summarized in Table~\ref
{tab:m87_ssp_snr}.

\begin{table}
\centering
\caption{Median SNR of SSPs detected in different parts of the jet and for
different scales.}
\label{tab:m87_ssp_snr}
\begin{tabular}{cccc}
\toprule

Region & Scale 1 & Scale 2 & Scale 3 \\
\midrule
Limbs, $z_{obs} \le 4 $ mas  & $17.2$ & $20.6$ & $61.9$ \\
Limbs, $z_{obs} > 4 $ mas  & $6.4$  & $7.6$  & $11.2$ \\
Center, $z_{obs} \le 4 $ mas & $7.3$  & $9.2$  & \textrm{-} \\
Center, $z_{obs} > 4 $ mas & $6.8$  & $7.4$  & $5.0$ \\

\bottomrule
\end{tabular}
\end{table}

Figure~\ref{fig:m87_wise_decomposition} shows an example of the SWD/IWD
decomposition of the VLBA image of M\,87 from the epoch  2007/02/17. The
internal structure of the flow is well sampled at Scales 1 and 2 of the SWD,
while the SWD Scales 3 and 4 are essential for providing an effective reference
frameset for the successful determination of the SSP displacement made using the
MCC algorithm.

The MCC algorithm has been applied to each consecutive pair of the jet images.
Based on the previously reported measurements made at these distances in the jet
\citep{reid_vlbispeeds_1989,biretta_detection_1995,dodson_propermotion_2006,
kovalev_inner_2007}, one can expect the jet to manifest a mixture of large and small
displacements. To ensure robust detectability of the entire range of potential
displacements, we use a tolerance factor of 1.5 and a correlation threshold of
0.6 in our MCC analysis. Finally, velocity constraints are set according to the
expected largest displacement ranges: on the jet side, the longitudinal
velocities are allowed to range between $-$2\,mas/yr and $+$18\,mas/yr,
transversally they are allowed to range between $-$5\,mas/yr and $+$5\,mas/yr.
On the counter-jet side, velocities ranging between $-2$ mas/yr and 2 mas/yr
longitudinally and transversally are allowed. For the analysis of the fine 
two-dimensional velocity structure of the jet, only the displacements determined at
the lowest SWD scale (Scale 1: 0.07 mas) are considered. The displacements
determined at coarser scales were used in the MCC and overall are consistent
with the ones obtained at the 0.07 mas scale.

\begin{figure}
    \centering
    \includegraphics{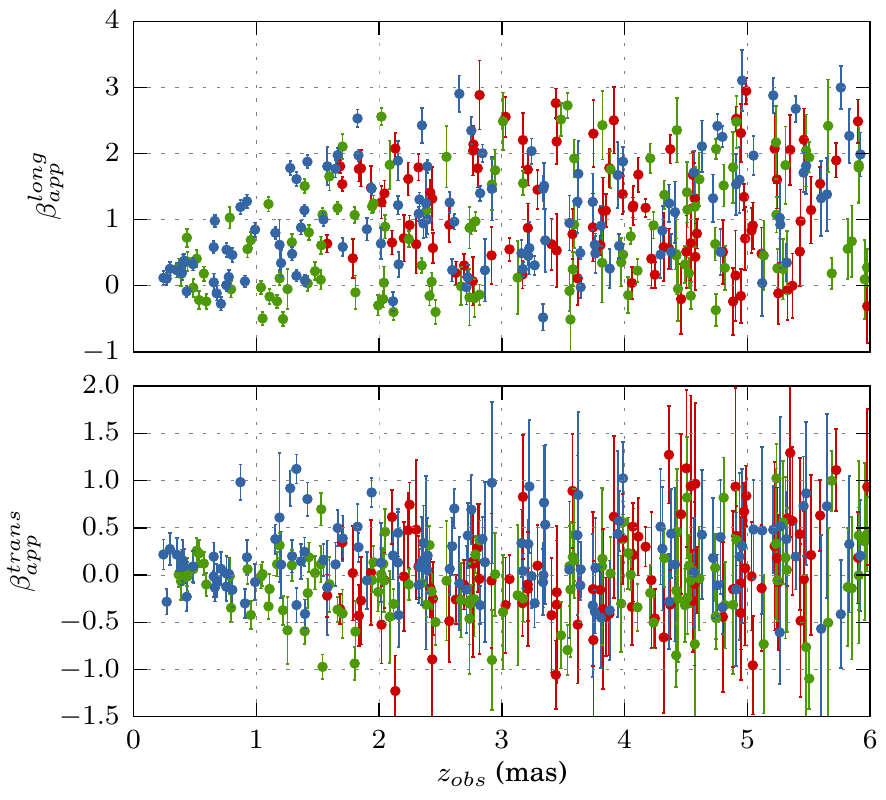}
    \caption[Longitudinal and transverse velocities of SSP measured in
    the jet of M\,87.]
    {Longitudinal (top) and transverse
      (bottom) velocities of significant structural patterns
      identified in the M\,87 jet. Colors indicate respective jet
      regions in which given SSPs are located (following the color
      coding introduced Fig.~\ref{fig:m87_velocity_field}): green and
      blue mark the SSPs from the northern and southern limb,
      respectively, and red marks the SSPs from the central stream.}
    \label{fig:m87_ind_velocities}
\end{figure}

The SWD yields a total of 68 SSPs that are detected on average in each
single image of the jet. At the smallest spatial scale of the SWD,
SSP matching is successfully made for 51\,\% of the SSPs, resulting in
detection of a total of 391 displacement vectors. The normalized cross
correlation between matching sets of SSPs is larger than 0.8, on
average. The errors of the displacements and velocity vectors are
obtained from the uncertainties of the SSP positions which are
computed using the stacked cross-correlation of the
SSP~\citep{mertens_detection_2016}, which yields:
\begin{equation}
\label{eq:feature_pos_error}
\sigma_x = \frac{b_x}{\sqrt{2}\ \mathrm{SNR}},\ \sigma_y = \frac{b_y}{\sqrt{2}\ \mathrm{SNR}}\,, 
\end{equation}
where $b_x$ and $b_y$ describe the beam size along the $x$
and $y$ coordinate, respectively, and $\mathrm{SNR}$ is the
signal-to-noise ratio of the SSP detection.

\subsection{SSP displacements and jet kinematics}
\label{sc:jetkinem}

The SSP displacements obtained using this approach are shown in
Fig.~\ref{fig:m87_velocity_field}, and the transverse and longitudinal
components of the corresponding velocities are plotted in
Fig.~\ref{fig:m87_ind_velocities}. 


Figures~\ref{fig:m87_velocity_field}--\ref{fig:m87_ind_velocities}   indicate
that the jet exhibits three kinematically distinct regions of organized 
stream-like features with predominantly   longitudinal displacement vectors. These
features will hereafter be   referred to as the northern and the southern limbs
and the central   ``stream'' (we prefer not to use the term ``spine'' for this
feature   as it is not clear whether it is indeed related to the physical
relativistic spine of the stratified outflow in M\,87).

The kinematics is more complex in the central stream region which
manifests both outward motions and substantial transverse
displacements.  The displacements can be reliably detected in the
central stream at $z_{\mathrm{obs}} \gtrsim 1.75$\,mas, when the width
of the jet exceeds three beam sizes.

The top panel of Figure~\ref{fig:m87_ind_velocities} shows the presence of both
slow (subluminal) and fast (superluminal) speeds in the central stream and each
of the limbs. This spread of apparent speeds might indicate jet stratification.
In each of the three regions, we also detect several outwards motions which
are most
of the time compatible with no motion (taking into account position uncertainty)
but might also be related to instability pattern motion.

An apparent acceleration can be seen in both limbs over the innermost 2\,mas of
the jet. At a distance $z_{\mathrm{obs}} \sim 0.25$\,mas, the maximum apparent
velocity is $\beta_{\mathrm{app}} \sim 0.2\,c$, which then linearly increases up
to $\beta_{\mathrm{app}} \sim 2.5\,c$ at $z_{\mathrm{obs}} \sim 2$\,mas. Between
$z_{\mathrm{obs}} \sim 2$\,mas and $z_{\mathrm{obs}} \sim 6$\,mas, we measure
either a slow acceleration or no acceleration at all. In this region, a maximum
velocity of $\beta_{\mathrm{app}} \sim 3\,c$ is found at $z_{\mathrm{obs}} \sim
5$\,mas. We note that because this acceleration is observed at the limbs of
the jet, the measured apparent velocities are only marginally affected by the
jet opening angle (see Eq.~\ref{eq:edge_viewing_angle}), and a fast
decrease in opening angle between 0 and 2 mas would not be able to explain the
observed increase in apparent velocity as this decrease would strongly affect
only the velocities measured in the central stream, which is not detected at
distances smaller than $\approx 2$\, mas.

We also observe an oscillation of the speed measured along the
jet, with minima in the jet speed found at $z_{\mathrm{obs}} \sim$
1.2~mas, 2.25~mas and 4.25~mas.

On the counter-jet side, there is only one SSP component that shows
consistently outward motion over five subsequent epochs (orange arrows in
Fig.~\ref {fig:m87_velocity_field}). At this distance from the core, the
counter-jet also appears limb-brightened, and the apparent position of this
feature is consistent with it being located in the northern limb. We can use it's
speed and brightness relative to the main jet to obtain estimates of the jet
angle to the line of sight and intrinsic speed.

Several other features in the outer part of the counter-jet exhibit slower
motions tending toward the northward direction, which may be viewed as a hint
for the clockwise rotation, as viewed by the observer, of the emitting
material.
The robustness of detection of the displacements of these features and potential
physical interpretation of these displacements should be a matter of further,
more detailed investigation.

In order to assess the robustness of the SSP and displacement identification, we
perform a bootstrapping analysis by randomly shuffling the individual images and
carrying out the WISE analysis on a number of such random sets. We found that on
average only 35\% of the SSPs can be matched in the pairs of shuffled images and
only 13\% of those can be traced over five or more epochs, compared to 24\% for
the non-shuffled set of images. The bootstrapping datasets also do not produce
regular velocity patterns and do not show the correlation between
$z_{\mathrm{obs}}$ and $\beta_{long}$ observed in the original sequence of
images. This demonstrates that the solution found by WISE does not result from a
spurious correlation.


\subsection{Flow stratification analysis}
\label{sc:m87:scc}


The large variations of velocity measured along and across the jet indicate the
complex physical nature of the flow, with both acceleration and stratification
of the flow likely to play important roles. To investigate the potential effect
of flow stratification on the observed velocity field we combine the WISE
velocity field analysis with the stacked cross correlation (SCC)
analysis~\citep{mertens_detection_2016}. The SCC method combines the  
cross-correlation results for all SSPs identified at different scales of the SWD in
all different pairs of epochs. This procedure yields a two-dimensional
distribution of the cumulative correlation coefficients, with peaks
corresponding to different velocity components. The significance and uncertainty
of each of these peaks is evaluated using Monte Carlo simulations.

For the present analysis, the SSPs detected at the three lowest SWD scales are used.
The jet is divided in two different regions (region 'A' at 0.5--1.0
mas distance from the core and region 'B' at 1--4 mas distance from
the core), and the SCC analysis is performed independently for each of
these regions. The extent of these two regions is constrained by the
requirement for the detection of statistically significant velocity
components in the SCC analysis of both the limbs and the central stream of
the flow.

\subsubsection{Region 'A'}

\begin{figure}
    \centering
    \includegraphics[width=\columnwidth]{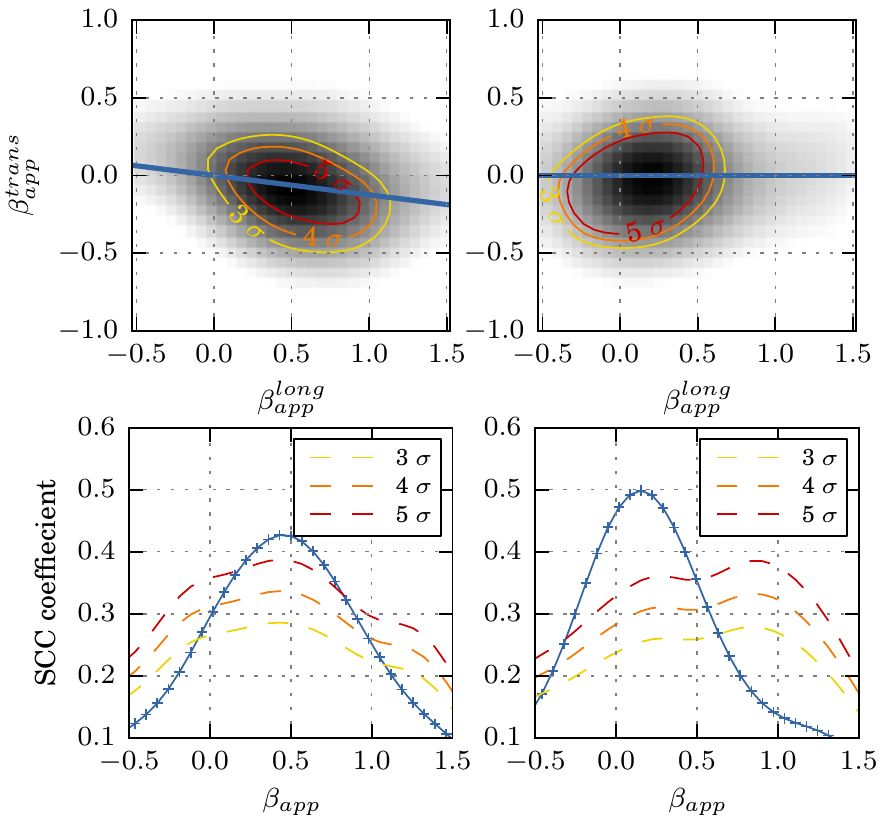}
    \caption[Results of the stacked cross correlation analysis of the
    northern and southern limbs of the Region 'A' in jet at distances
    between 0.5 and 1\,mas from the core.]
    {Results of the stacked cross correlation analysis of the northern
      (left panel) and southern limbs (right panel) in Region 'A'
      of the jet at distances between 0.5 and 1\,mas from the core.
      The velocity cross correlation map is shown at the top in
      gray scale. Contours indicate the statistical significance of the
      correlation. A slice along the blue line is shown in the bottom
      panels. A single velocity component with more than $5\,\sigma$
      significance is found in both limbs. The parameters for this
      velocity component are listed in
      Table~\ref{tab:m87_gncc_full_05_1mas}.}
    \label{fig:m87_gncc_full_05_1mas}
\end{figure}

In the northern part of the jet, the SCC is computed from a total of
42 SSPs. The resulting velocity cross correlation maps and the
corresponding significance estimates are plotted in
Fig.~\ref{fig:m87_gncc_full_05_1mas}. For the northern limb of
region 'A', the SCC results imply a single velocity component, with an
apparent speed of $\beta^{\mathrm{north}}_{\mathrm{app}} = 0.48 \pm
0.06 c$ and a significance of $5.9\,\sigma$. In the southern limb, the
SCC has been applied to a total of 47 SSPs, also revealing a single
velocity component with $\beta^{\mathrm{south}}_{\mathrm{app}} = 0.21
\pm 0.04 c$ and a significance of $8.7\,\sigma$.  The results are
further summarized in Table~\ref{tab:m87_gncc_full_05_1mas}.

The SCC analysis did not unveil any stratification close to the
core. This result confirms however the speeds measured in
Sect.~\ref{sc:m87:wise_velocity_7mm}, further strengthening the
conclusion that subluminal speeds are generally found in this
region. If any stratification would be present, the respective
velocity difference must have been too small so
that it would remain undetected with the SCC.  In this case, the
apparent velocity measured by this method can be interpreted as a
weighted mean velocity of the different jet layers along a given line of
sight. This may also explain the difference between the speeds
measured in the northern and southern limb (an alternative explanation
may involve rotation of the flow, which will be examined later).

\begin{table}
\centering
\caption{Velocity components identified from stacked cross correlation analysis
  of the northern and southern limbs in Region 'A', at distances between 0.5 and 1\,mas from the core.
}
\label{tab:m87_gncc_full_05_1mas}
\begin{tabular}{ccccc}
\toprule
 Region A & $\beta_{\mathrm{app}}^{trans}$ & $\beta_{\mathrm{app}}^{long}$ &
 $\beta_{\mathrm{app}}$ &
 $\sigma$ \\
\midrule
\textit{North} & $-0.10 \pm 0.03$  & $0.47 \pm 0.05$ & $0.48 \pm 0.06$ & 5.9 \\
\textit{South} & $-0.02 \pm 0.03$  & $0.21 \pm 0.03$ & $0.21\pm 0.04$ & 8.7 \\
\bottomrule
\end{tabular}
\begin{tablenotes}
  \small
  \item \textbf{Note:} $\sigma$ -- statistical significance of the SCC peak;
\end{tablenotes}
\end{table}

\begin{figure*}
    \centering
    \includegraphics{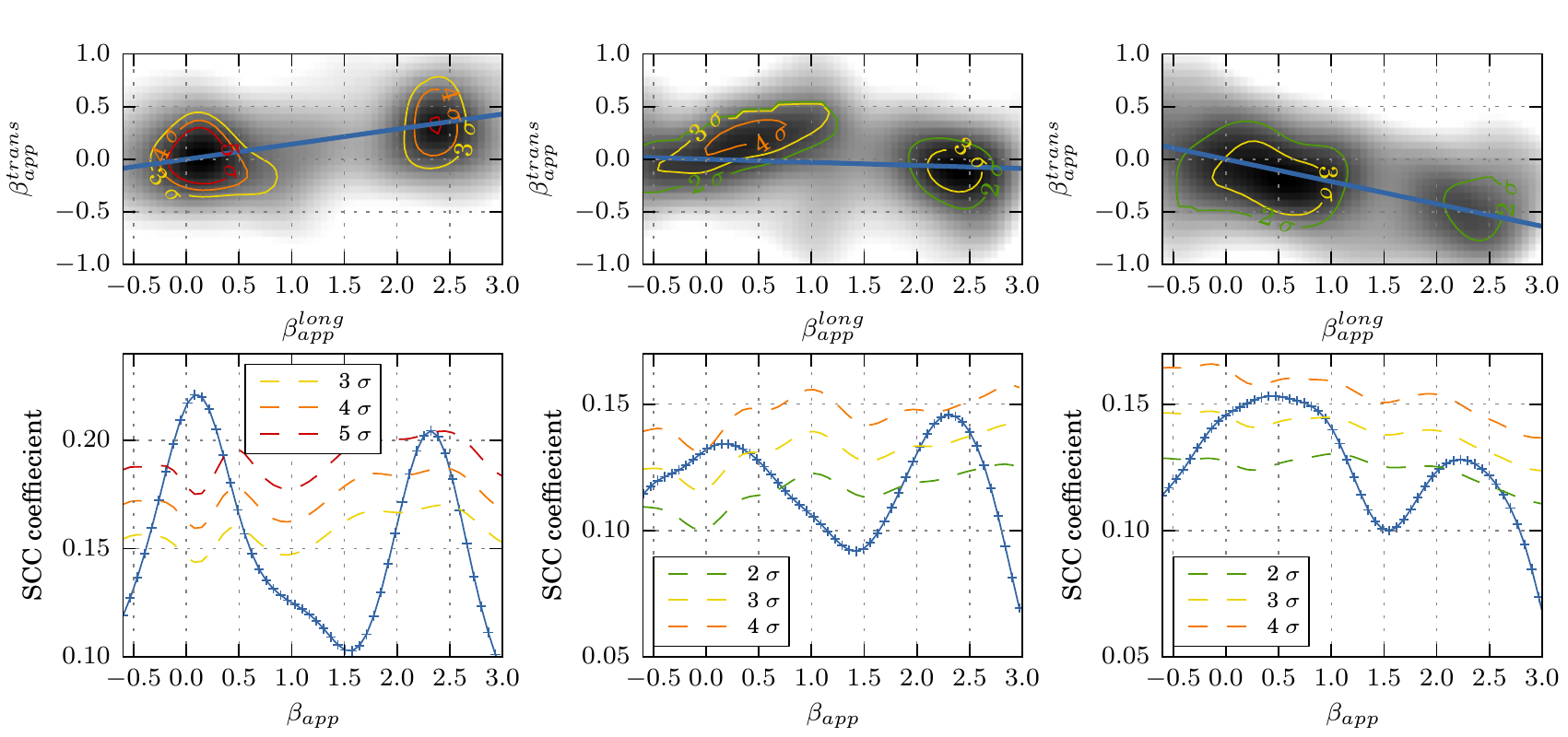}
    \caption[Results of the stacked cross correlation analysis of the
    northern and southern limbs and the central stream in Region
    'B' of the jet at distances between 1 and 3\,mas from the core.]
    {Results of the stacked cross correlation analysis of the northern
      limb (right panel), central stream (middle panel), and southern
      limb (left panel) in Region 'B' of the jet, at a distance
      between 1 and 4\,mas from the core.  The cross correlation map
      is shown at the top in gray scale along with contours of the
      significance. A slice along the blue line is shown at the
      bottom. Two main velocity components are consistently found in
      the three regions of the jet. The derived parameters for the
      velocity components are listed in
      Table~\ref{tab:m87_gncc_full_1_4mas}.}
    \label{fig:m87_gncc_full_1_4mas}
\end{figure*}

\subsubsection{Region 'B'}

For region 'B' (1--4 mas from the core), the SCC analysis was
performed independently for each of the limbs and for the central
stream of the jet, using 245, 267, and 217 SSPs, respectively. The
resulting cross-correlation maps are plotted in
Fig.~\ref{fig:m87_gncc_full_1_4mas}. The measured correlations reveal the
presence of two distinct velocity components, which suggest a strong
stratification of the flow with a slow, mildly relativistic speed
($\beta_{\mathrm{app}} \sim 0.35\,c$) and a faster, relativistic speed
($\beta_{\mathrm{app}} \sim 2.3\,c$). The velocity parameters derived
for these components are summarized in
Table~\ref{tab:m87_gncc_full_1_4mas}.

The statistical significance of most of the velocity components found
in region 'B' is better than $\approx 4\,\sigma$. The overall
consistency of the speeds found in the limbs and the central stream lends
further support for stratification of the flow. A somewhat low ($2.5 \sigma$) formal
significance of the correlation is found for the fast velocity
component in the southern limb of the jet. This may result from a
broader range of measured displacement velocities registered there as
compared to the northern limb.

\begin{table}
\centering
\caption{Velocity components identified using the stacked cross correlation
analysis
in Region 'B', at 1--4\,mas distance from the core.}
\label{tab:m87_gncc_full_1_4mas}
\begin{tabular}{ccccc}

\toprule
Region B & $\beta_{\mathrm{app}}^{trans}$ & $\beta_{\mathrm{app}}^{long}$ &
$\beta_{\mathrm{app}}$ & $\sigma$
 \\
\midrule

\multicolumn{4}{l}{\text{Fast velocity component}}\\
\textit{North} & $0.27 \pm 0.04$  & $2.40 \pm 0.03$ & $2.41 \pm 0.05$ & 5.1 \\
\textit{Center} & $-0.1 \pm 0.1$   & $2.32 \pm 0.17$ & $2.32 \pm 0.20$ & 4.0 \\
\textit{South} & $-0.41 \pm 0.07$ & $2.16 \pm 0.14$ & $2.20 \pm 0.15$  & 2.5 \\

\multicolumn{4}{l}{\text{Slow velocity component}}\\
\textit{North} & $0.01 \pm 0.03$  & $0.17 \pm 0.04$ & $0.17 \pm 0.06$ & 8.0 \\
\textit{Center} & $0.13 \pm 0.12$  & $0.34 \pm 0.30$ & $0.37 \pm 0.32$ & 4.4 \\
\textit{South} & $-0.14 \pm 0.07$ & $0.47 \pm 0.22$ & $0.49 \pm 0.24$ & 3.9 \\
\bottomrule
\end{tabular}
\begin{tablenotes}
  \small
  \item \textbf{Note:} $\sigma$ -- statistical significance of the SCC peak;
\end{tablenotes}
\end{table}

\subsection{Jet collimation}
\label{sc:jet_collimation}

In the magnetically accelerated jet model, the conversion from
Poynting flux into kinetic energy depends on the configuration of the
magnetic field
lines~\citep{begelman_asymptotic_1994,lyubarsky_transformation_2010}. The
jet collimation profile is therefore an important indicator of the jet
acceleration mechanism. The segmented wavelet decomposition (SWD)
method introduced in~\cite{mertens_waveletbased_2015} provides a
reliable way to analyze the evolution of the jet width (jet radius)
with distance from the core. 

To determine the jet collimation profile, three stacked-epoch wavelet scale
(SWS) images of M\,87 are prepared. Each of these images comprises all SSPs
detected at a given scale $j = 1, 2, 3$ at all 11 epochs. Near to the jet base,
the jet limb is well described by the SWS at the smallest wavelet scale. As the
jet expands, the width of each of its limbs is best described by progressively
larger SWS scales. Scale 1 (0.07 mas) of the SWS is used between
$z_{\mathrm{obs}} = 0.4$~mas and $z_{\mathrm{obs}} = 0.8$~mas, scale 2
(0.14 mas) is then used up to $z_{\mathrm{obs}} = 1.6$~mas,   and finally
scale 3 (0.21 mas) is selected at larger core   separations.

\begin{figure}
    \centering
    \includegraphics{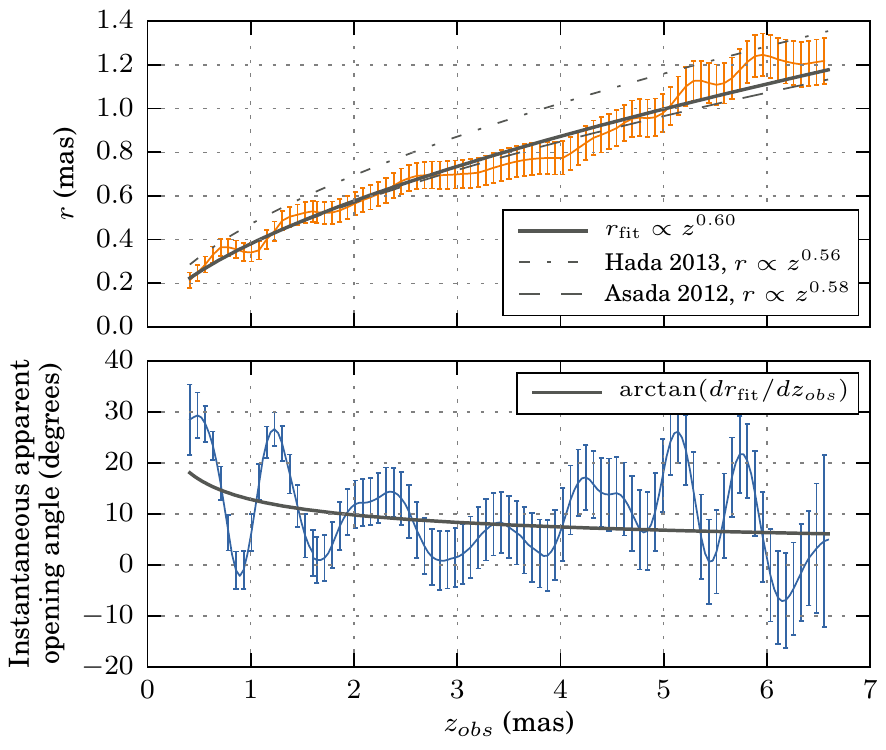}
    \caption[Jet radius and apparent opening angle of the
    jet in M\,87 with distance from the core.]
    {\label{fig:m87_jet_radius_opening_angle} Evolution of the radius
      (top panel) and instantaneous apparent opening angle (bottom
      panel) of the jet in M\,87 with distance from the core. A
      power law fit to the jet radius is consistent with results
      obtained by \cite{asada_structure_2012} (dashed line) and
      \cite{hada_innermost_2013} (dashed dotted line). Evolution
      of the instantaneous opening angle corresponding to the power
      law fit obtained for the jet radius is shown by the solid
      gray line in the bottom panel.}
\end{figure}

The jet radius is estimated by taking transverse profiles of the
stacked-epoch SWS images of the jet at
regular intervals ($\delta z = 1~\mathrm{px} = 0.035$~mas) and
measuring the distance between the two limbs of the jet. A Gaussian
profile is fitted to each of the limbs to further improve this
calculation. It is particularly difficult to obtain a robust
uncertainty estimate for these measurements. Our conservative approach
is to assume that the position error of limb determination is 1/4 of
the beam (0.07 mas) at the core of the jet, and linearly increases
to reach 1/2 of the beam at the last profile, located at 6.6 mas from
the core. The result is plotted in
Fig.~\ref{fig:m87_jet_radius_opening_angle}. The instantaneous
apparent jet opening angle, $\Theta_\mathrm{app}$, is computed from
the measured jet radii using the relation
$\Theta_{\mathrm{app}} = \arctan(dr / dz_{\mathrm{obs}})$.
The local curvature $dr / dz_{\mathrm{obs}}$ is approximated from a linear fit
of the jet radius 0.2 mas before and after each point of the calculation.

The jet collimation profile is well described by a power law $r
\propto z^k $ with $k = 0.60 \pm 0.02$ (reduced $\chi^2 = 0.55$). This
result is consistent with previous measurements made by
\cite{asada_structure_2012} ($k = 0.58 \pm 0.02$) and
\cite{hada_innermost_2013} ($k = 0.56 \pm 0.03$). As seen from
Fig.~\ref{fig:m87_jet_radius_opening_angle}, expansion of the jet
reveals an oscillatory pattern, most likely reflecting repeated
over-collimation and over-expansion of the flow. The over-collimation
is visible at $z_{\mathrm{obs}} \sim$ 1, 1.9 and 4~mas, and the
over-expansion manifests itself at $z_{\mathrm{obs}} \sim$ 0.75, 1.4
and 2.6~mas, suggesting that the spatial period of these oscillations
increases with the distance as $\sim z_{\mathrm{obs}}$. Oscillations
of the jet width seen at distances $\gtrsim 5$ mas have a shorter
period of $\approx 0.5$ mas.

It is interesting to note that oscillations of the jet width
correlate with the changes in the apparent jet speed reported in
Sect.~\ref{sc:jetkinem}.  The locations of maximum over-collimation of
the jet correspond well to the observed minima in the apparent speed
of the jet. If these oscillations reflect the hydrodynamics of the
flow, the pressure ratio between the jet and the external medium might
be close to unity~\citep{daly_gasdynamics_1988}. Presence of
oscillating patterns in an expanding flow is also expected for an
expanding magnetized
jet~\citep{lyubarsky_asymptotic_2009,komissarov_stationary_2015}. Alternatively, these
oscillations might reflect the evolution of the pinch mode or the
elliptical surface mode of K-H
instability~\citep{hardee_3dkh_2000,lobanov_khmodel_2001}.

\section{WISE analysis of 2 cm VLBA observations}
\label{sc:m87:wise_velocity_2cm}

Between 1995 and 2010, the jet in M\,87 was regularly observed as part of the
MOJAVE project \citep{lister_mojave_2009,lister_mojave_2013}. The MOJAVE
database contains 30 images of this jet, providing, on average, one observation
every seven months. With this cadence, WISE can reliably detect jet speeds of up
to $\beta_{\mathrm{app}} \sim 2\,c$, assuming a maximum detectable displacement
of 4 times the beam size. In an analysis of this data using the Gaussian 
model-fitting technique,~\cite{kovalev_inner_2007} reported only a slow, subluminal
motion with a maximum speed $\beta_{\mathrm{app}} \sim 0.05$\,c. Finding this
result to contradict several physical aspects of the jet, including the 
counter-jet flux density ratio, \cite{kovalev_inner_2007} suggested that the measured
subluminal speed might represent pattern motions of either shocks or MHD
instabilities. Using the WISE analysis, we can attempt to detect a faster
speed
from the MOJAVE data. The   application of the SWD made for this purpose is
performed using four   scales, ranging from 0.4 mas (scale 1) to 3.2 mas (scale
4).

The interval between two epochs is generally too long to warrant
detection of the fast velocity $\beta_{\mathrm{app}} \sim 2.5$\,c that
has been measured in the 43\,GHz VLBA data. We concentrate therefore
on the pair of observations separated by the shortest time
interval. The complete MOJAVE database has two epochs (2000/04/07 and
2000/06/27) that matched this criteria, and we complement them with
two full track 15\,GHz VLBA observations obtained at epochs 2000/01/22
and 2000/05/08~\citep{kovalev_inner_2007}. 

\begin{figure}
    \centering
    \includegraphics{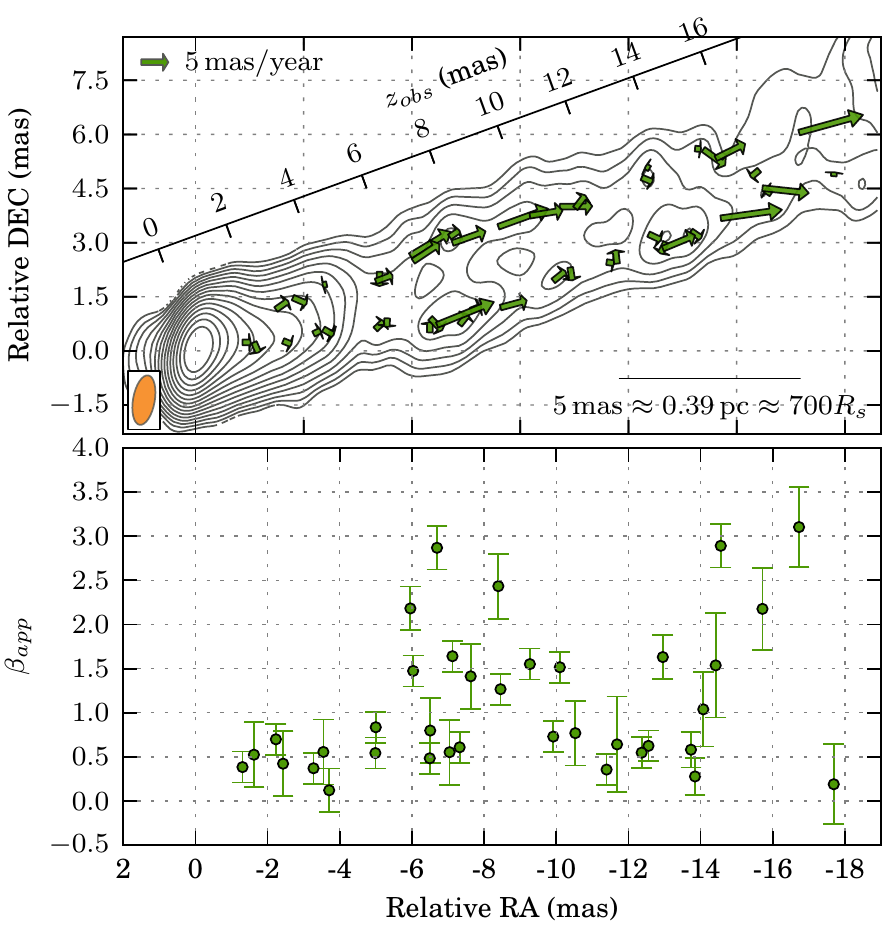}
    \caption{Two-dimensional velocity field traced by the
      structural patterns detected using the WISE analysis of four
      epochs of 15\,GHz VLBA observations of M\,87 made between
      2000/01/07 and 2000/06/27 (upper panel). The tracks are
      overplotted on a stacked-epoch image of the jet. The
      corresponding apparent velocities are plotted in the bottom
      panel.}
\label{fig:velocities_map_stack_mojave}
\end{figure}

The result of applying the WISE algorithm to these four VLBA images at
15\,GHz is plotted in Fig.~\ref{fig:velocities_map_stack_mojave}. We
detected fast relativistic speeds of $\beta_{\mathrm{\mathrm{app}}}
\sim 2$--$3\,c$ for 6 different SSPs in both the northern and
southern limbs of the jet at $z_{\mathrm{obs}} \sim 6.5$\,mas and
$z_{\mathrm{obs}} \sim 16$\,mas. In addition to this, 8 SSPs exhibit apparent
superluminal speeds of $\beta_{\mathrm{\mathrm{app}}} \sim 1.5\,c$, and
subluminal speeds ($\beta_{\mathrm{\mathrm{app}}} \sim 0.5\,c$).

The displacements traced in the 15 GHz VLBA images do not provide
enough material for tracing the velocity field in the jet. However,
one can see that full-tracks VLBA observations at 15\,GHz made at
$\sim 2$ month intervals should be sufficient to obtain a reliable
velocity field. 

\section{Discussion}
\label{sc:discussion}

\subsection{Jet internal structure}
\label{sc:jet_stratification_model}

\begin{figure}
    \centering
    \includegraphics{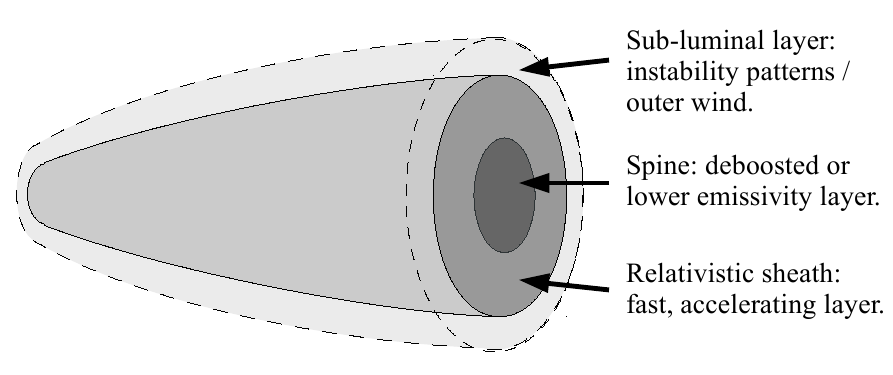}
    \caption{\label{fig:jet_stratification_model} Schematic
      illustration of the internal structure of the M\,87 jet adopted
      in the analysis presented in this paper. The superluminal
      kinematic feature in the jet is associated with the sheath and
      the subluminal feature is related to a slower, outer layer of the
      flow. The flow is also expected to feature a fast inner spine,
      which however cannot be detected in the present data either
      because of its lower emissivity compared to the emissivity of
      the sheath layer or because its speed is too high to be detected
      with the present data.}
\end{figure}

The stacked cross correlation analysis discussed in
Sect.~\ref{sc:m87:scc} reveals a clear stratification of the flow,
with a slow, subluminal component and a faster relativistic component
detected in all three streams inside the jet. 
Both the slow and the fast velocity components exhibit similar speeds
in each of the streams, which suggests that each of them reflects plasma
motions occurring at similar radial separations (similar jet layers)
in a transversally stratified flow.
Since the fast velocity component is present in all three streams, it
can be associated with the sheath, rather than the spine, of the
jet. A faster spine is either not present in the flow (at least at the
linear scales studied here), or it is deboosted and appears dimmer than
the slower moving material of the sheath (in this case, VLBI
observations at time intervals of smaller than 10 days are required to
successfully disentangle this faster component from the two velocity
components identified in this paper). In further analysis, we assume
the latter possibility, which corresponds to the schematic view of the
jet shown in Fig.~\ref{fig:jet_stratification_model}.

The observed jet stratification, with a fast, accelerating sheath
and a subluminal layer can be interpreted in several different
ways. It may reflect a 'two-fluid' jet \citep{sol_twoflow_1989}
comprising a relativistic beam (or spine) originating either from the
magnetosphere of the central engine, or from the inner part of the
accretion disk, and a non-relativistic wind originating from outer
parts of the accretion disk \citep{tsinganos_magnetic_2002}.  The
viability of this scenario has been tested in RMHD simulations of AGN
jets \citep[e.g.][]{komissarov_magnetic_2007} which demonstrated that
magnetic field lines originating from different parts of the accretion
disk would result in substantial transverse stratification of the
flow.

Alternatively, the fast sheath component may reflect the true speed of
the flow (without substantial transverse stratification), while the
slower one would then be interpreted as a pattern speed associated
with current driven (CD) or Kelvin-Helmholtz (K-H) instability. In
general, K-H instability pattern speed is expected to be considerably
slower than the flow speed \citep{hardee_3dkh_2000}. It was also
suggested that the subluminal apparent speeds of $\sim 0.5\,c$
detected in the jet in M\,87 at kiloparsec scale indicate a pattern
speed from K-H instability \citep{lobanov_internal_2003}.

\subsection{Counter-jet and jet viewing angle}
\label{sc:m87:viewing_angle}

The presence of a counter-jet in M\,87 jet has been suggested based on
the structure detected in 15\,GHz VLBA
maps~\citep{kovalev_inner_2007}. A similar structural detail
resembling a counter-jet is also observed in the 43\,GHz images of
M\,87~\citep{ly_high_2007,walker_vlba_2008}. The WISE analysis
performed on the 43\,GHz VLBA data has revealed several SSPs located in
the counter-jet structure and moving outward with speeds
similar to the subluminal speed measured on the jet side at the same
distances from the core. We will use this velocity measurement to
constrain the viewing angle of the jet.

Relativistic Doppler beaming affects differently the flux density in the jet
and the counter-jet. If the jet is seen at a viewing angle $\theta$, the
observed intensity ratio between the jet and counter-jet is given by:
\begin{equation}
\label{eq:intensity_ratio}
\mathcal{R} = \frac{I_{\mathrm{jet}}}{I_{\mathrm{cjet}}} = \left[\frac{\gamma_{cj} (1 +
\beta_ {cj} \cos
(\theta))}{\gamma_{j} (1 - \beta_{j} \cos(\theta))}\right]^{2 - \alpha}
\end{equation}
with $\beta_{j}$, $\beta_{cj}$ are the intrinsic speeds of the plasma
in the jet and counter-jet respectively, $\gamma_{j}$, $\gamma_{cj}$
are the corresponding Lorentz factors and $\alpha$ is the spectral
index of the synchrotron radiation, $I_\nu \propto \nu^\alpha$.  In a
recent publication, $\alpha = -1$ was estimated for the radio emission
on sub-parsec scales in the M\,87 jet \citep{hovatta_mojave_2014}.

\begin{figure}
    \centering
    \includegraphics{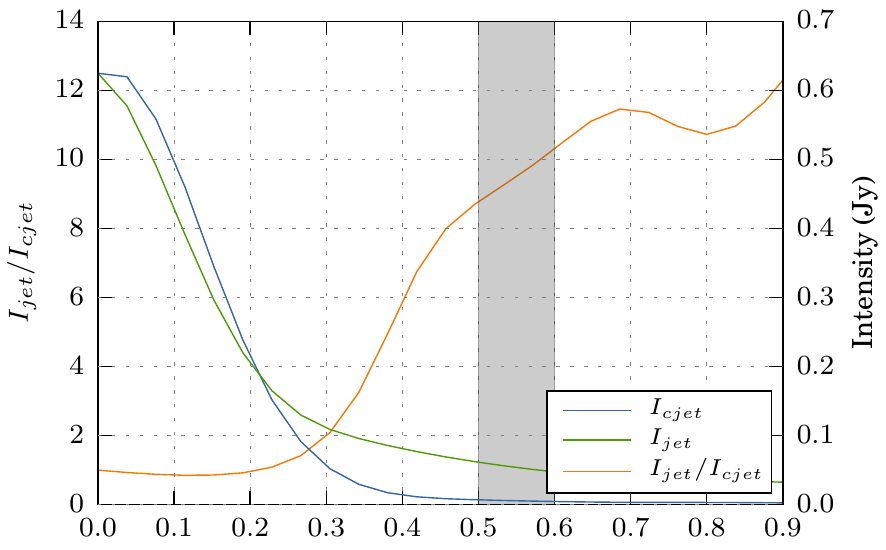}
    \caption[Jet and counter jet intensity measured on a longitudinal
    slice of the jet and corresponding ratio.]  {Jet (green) and
      counter-jet (blue) longitudinal intensity profile and their
      corresponding ratio (orange). The gray shading indicates the
      span travelled by the SSP detected on the counter-jet side for
      which the speed has been robustly estimated. In order to exclude
      biasing by the bright core of the jet, only measurements at core
      separations larger than 0.4\,mas ($\approx 1.4$ FWHM of the
      restoring beam) are considered in the analysis of the
      jet/counter-jet ratio.}
    \label{fig:jet_counterjet_ratio}
\end{figure}

We measure the intensity ratio $\mathcal{R}$ by taking a longitudinal profile of
emission brightness along the jet limbs. In order to reference properly the jet
and counter-jet locations, special attention must be paid to the identification
of the position of the core. Because of synchrotron self-absorption, the
observed location of the jet core does not correspond exactly to the true base
of the jet \citep{lobanov_ultracompact_1998}. However, in the case of M\,87,
\cite{hada_origin_2011} found that the central engine of M\,87 and the radio
core at 43 GHz are separated by a projected distance of only $41 \pm
12~\micro\mathrm{as}$, which is small enough to consider, in our case, the core
as the point of symmetry of the jet. The counter-jet component which yields a
robust speed estimate is located in the northern limb of the counter jet. To
preserve the symmetry, the jet to counter-jet intensity ratio is measured
between the northern limb of the counter jet and the southern limb of the jet.
To obtain the profile, flux density is measured at every pixel (0.035 mas; 1/8
of the beam size). The result is shown in Fig.~\ref{fig:jet_counterjet_ratio}.
We note here again that both the intensity ratio and velocities are measured
at the limbs of the jet, and therefore the effect of variable jet opening angle
near the core would only be marginal and well below the statistical
uncertainties of our measurements.

\begin{figure}
    \centering
    \includegraphics{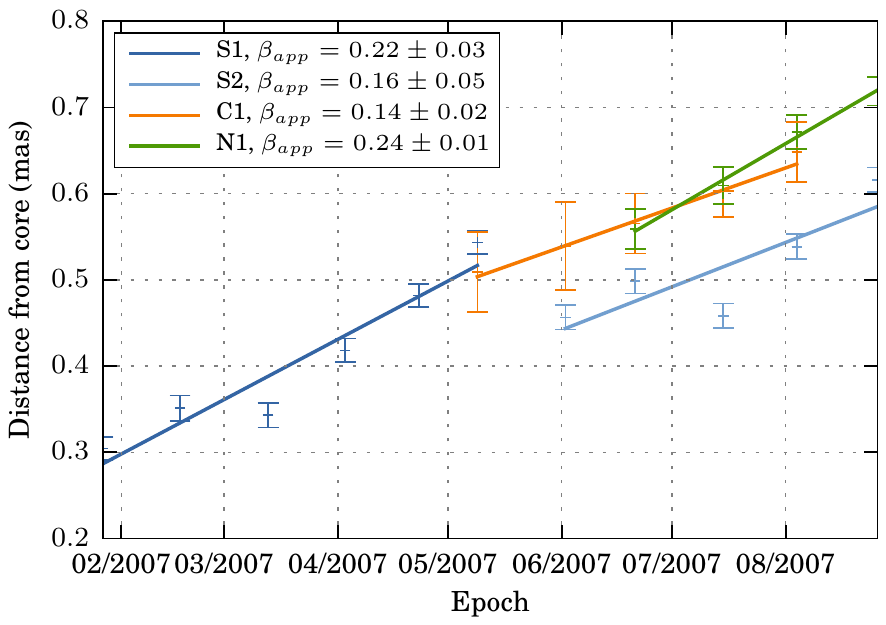}
    \caption[Core separations of the SSP that we tracked for at
    least 4 epochs over a 0.8 mas extent in both the jet and counter-jet
    direction.]
    {Core separations of the SSPs that were tracked for at
    least 4 epochs over a 0.8 mas extent in both the jet and counter-jet
    direction. The jet components S1 (dark blue) and S2 (light blue)
    are detected in the southern limb of the jet, while the jet component N1
    (green) is
    detected in the northern limb of the jet. On the counter jet side, one
    component, C1 (orange) is detected in the northern limb of the flow.}
    \label{fig:m87_dfc_0-08mas}
\end{figure}

In order to avoid biasing by the bright core of the jet we exclude the
inner 0.4~mas portion of the flow (corresponding to $\approx 1.4$ FWHM
of the restoring beam).  At separations between 0.4~mas and 0.8~mas,
we observe a sharp increase of the jet to counter-jet intensity ratio
from $\sim 7$ to $\sim 11$ which may result from acceleration of
the flow. In this region, robust displacement detections have been
made in three components on the jet side (with speeds ranging from
$0.16\,c$ to $0.24\,c$) and in one feature on the counter-jet side
with an apparent speed of $0.14\,c$. These components are presented in
Fig.~\ref{fig:m87_dfc_0-08mas}.

\begin{figure}
    \centering
    \includegraphics{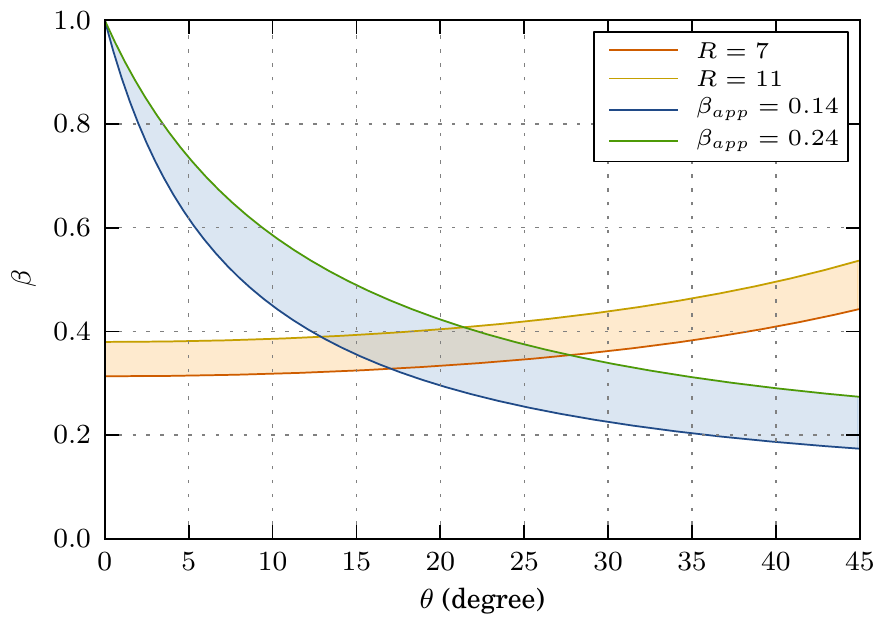}
    \caption{The jet viewing angle constraints obtained from the minimum and
    maximum jet
    to counter jet intensity ratio $R$ and apparent velocity $\beta_{\mathrm{app}}$
    measured at core separations between 0.4 and 0.8~mas.}
    \label{fig:m87_viewing_angle_beta_constaint}
\end{figure}

The combination of the observed jet to counter-jet intensity ratio and
the maximum and minimum measured apparent speeds provides conservative
constraints on the viewing angle of the jet. This is illustrated in
Fig.~\ref{fig:m87_viewing_angle_beta_constaint} which shows the ranges
of the viewing angle and the intrinsic speed that can reconcile both
the apparent speed and the measured jet to counter-jet intensity
ratio:
\begin{equation}
\label{eq:viewing_angle_counter_jet_large} 
13\degree \le \theta \le 27\degree,~0.32 \le \beta \le 0.4 \,.
\end{equation}

This estimate assumes that both the jet and the counter-jet have the
same intrinsic speed and brightness. In a stratified flow, it is
however very likely that different layers are dominating the emission
in the jet and the counter-jet sides.  To take this into
account, we consider the difference between the apparent speed found in the jet and
the speed measured in the counter jet. The counter-jet component C1 is
detected at distances of 0.5--0.6 mas from the core (see
Fig.~\ref{fig:m87_dfc_0-08mas}). In this region, we measure a jet to
counter-jet intensity ratio $\mathcal{R} = 9.5 \pm 1.5$. On the jet
side, we can use a more robust velocity measurement obtained from the
stacked cross correlation analysis discussed in
Sect.~\ref{sc:m87:scc}. Between 0.5 and 1~mas from the core, we
measure a jet apparent speed $\beta^{\mathrm{jet}}_{\mathrm{app}} = 0.21 \pm
0.04\,c$ in the southern limb of the jet. Then, using
Eq.~\ref{eq:intensity_ratio}, we obtain:
\begin{equation}
\label{eq:viewing_angle_counter_jet}
\theta = 17.2 \pm 3.3 \degree,~ \beta_{\mathrm{jet}} = 0.42 \pm 0.07,~ \beta_{\mathrm{cjet}} =
0.33 \pm 0.06
\end{equation}
This estimate of the viewing angle is consistent with the constraint
imposed by the maximum apparent speed $\beta_ {\mathrm {app}} \sim
6\,c$ measured at HST-1 which requires a viewing angle $\theta
\lesssim 19 \degree$.

\subsection{Jet rotation}
\label{sc:m87:jet_rotation}

\subsubsection{Evidence and estimation of the jet rotation}

The stacked cross correlation analysis has revealed substantial
difference between the speeds measured in the northern and southern
limbs of the jet (see Table~\ref{tab:m87_gncc_full_05_1mas} and
Table~\ref{tab:m87_gncc_full_1_4mas}). At core separations of 0.5--1.0
mas, the significance of this difference reaches $\sim 4\,\sigma$, while it
is above $\sim 1.5\,\sigma$ for both the fast and slow velocity
components detected between 1 and 4~mas.

\begin{figure}
    \centering
    \includegraphics[width=\columnwidth]{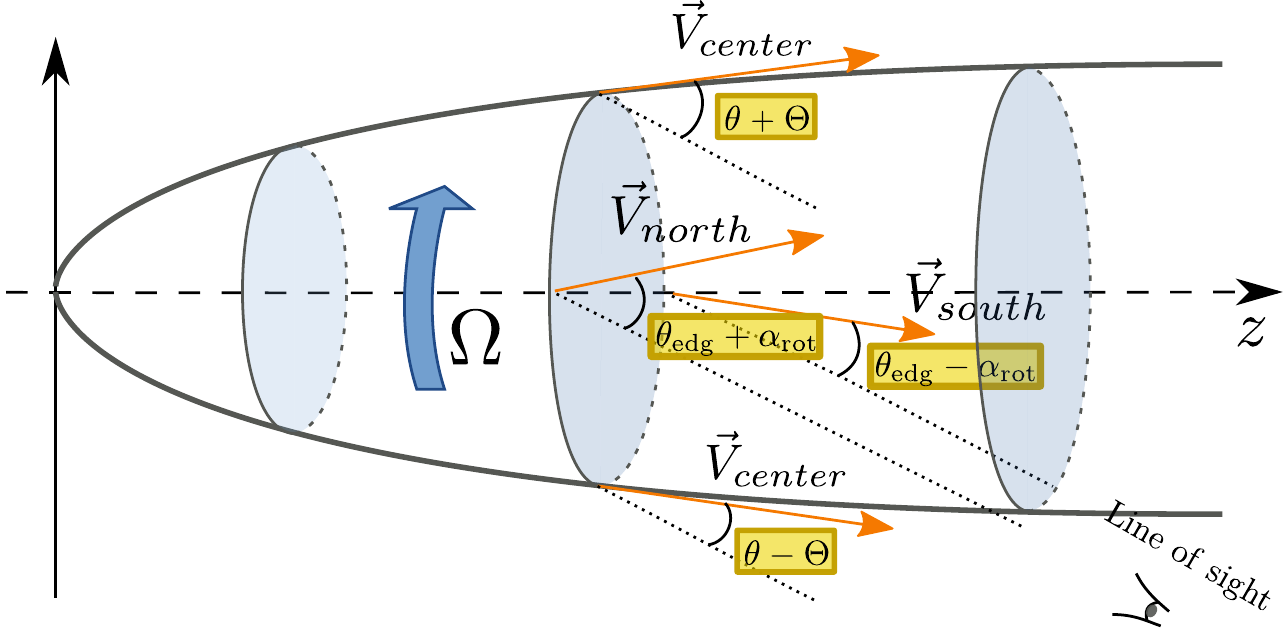}
    \caption[Schematic illustration of the effect of the jet rotation on the
    viewing angle in the three regions of the jet.]
    {\label{fig:jet_rotation_schema} Schematic illustration of the
      effect jet rotation has on the instantaneous viewing angles
      in the three jet streams. In a rotating outflow, the
      generic viewing angle of the jet axis, $\theta$ is modified by
      the angle, $\alpha_{\mathrm{rot}}$, between the instantaneous velocity vector
      and the jet axis, and by the jet intrinsic opening angle
      $\Theta$, as described by
      Eqs.~\ref{eq:effective_viewing_angle}--\ref{eq:edge_viewing_angle}.}
\end{figure}

This difference can be explained naturally by invoking jet rotation. The
intrinsic speed, $\beta$, of the flow can be decomposed into poloidal ($\beta_p
= \beta \cos\alpha_{\mathrm{rot}}$) and toroidal (or azimuthal, $\beta_ {\phi} =
\beta\,\sin\alpha_{\mathrm{rot}}$) components, where $\alpha_{\mathrm{rot}}$ is the angle between the
intrinsic velocity and the jet axis. In this framework, for a clockwise rotation
of the jet as viewed by the observer, the SSP components moving along the
northern limb of the jet would be viewed at a larger viewing angle, increased by
$\alpha_{\mathrm{rot}}$, while in the southern limb, the effective viewing angle would be
reduced by $\alpha_{\mathrm{rot}}$. This effect would not apply to the central stream of the
jet. However, since the velocities measured in the central stream can originate
either from the near-side or the far-side of the jet, the effective viewing
angle is increased by an angle $\psi$, which is zero when far-side and near-side
of the jet contribute equally to the velocity detected in the central stream,
and is bounded by the intrinsic instantaneous opening angle $\Theta$ of the jet.
For the velocity vectors measured in the limbs, one need also to take into
account the effect of the opening angle of the jet. The resulting geometrical
setting is illustrated in Fig.~\ref{fig:jet_rotation_schema}.

Following this scheme, the effective viewing angles for the three
streams in the jet can then be written as:
\begin{equation}
\label{eq:effective_viewing_angle}
  \begin{aligned} 
    &\theta_{\mathrm{north}} = \theta_{\mathrm{edge}} + \alpha_{\mathrm{rot}} \\
    &\theta_{\mathrm{south}} = \theta_{\mathrm{edge}} - \alpha_{\mathrm{rot}} \\
    &\theta_{\mathrm{center}} = \theta + \psi\,,\quad -\Theta \le \psi  \le \Theta\, ,
  \end{aligned} 
\end{equation}
where $\theta_{\mathrm{edge}}$ is the edge viewing angle corrected for
the opening angle of the jet:
\begin{equation}
\label{eq:edge_viewing_angle}
\cos\theta_{\mathrm{edge}} = \frac{\cos\theta}{(\tan^2\Theta + 1)^{1/2}} \,.
\end{equation}
The intrinsic opening angle $\Theta$ can be estimated from
the apparent opening angle measured in Fig.~\ref{fig:m87_jet_radius_opening_angle},
using $\Theta = \Theta_{\mathrm{app}} \sin\theta$.

Assuming that the flow has the same intrinsic velocity $\beta$ in the
three regions, the rotation angle $\alpha_{\mathrm{rot}}$ can be determined by
solving the system of two equations:
\begin{equation}
\label{eq:rotation_effect}
\begin{dcases}
\beta^{\mathrm{north}}_{\mathrm{app}} = \frac{\beta \sin(\theta_{\mathrm{edge}} + \alpha_{\mathrm{rot}})}{1 - \beta \cos(\theta_{\mathrm{edge}} + \alpha_{\mathrm{rot}})} \\
\beta^{\mathrm{south}}_{\mathrm{app}} = \frac{\beta \sin(\theta_{\mathrm{edge}} - \alpha_{\mathrm{rot}})}{1 - \beta \cos(\theta_{\mathrm{edge}} - \alpha_{\mathrm{rot}})}
\end{dcases} 
\end{equation}
If the velocity in the central stream is also considered, an additional equation
can be written:
\begin{equation}
\label{eq:rotation_effect_center}
\beta^{\mathrm{center}}_{\mathrm{app}} = \frac{\beta \sin(\theta + \psi)}{1 - \beta \cos(\theta + \psi)}
\end{equation}
A unique solution for
Eqs.~\ref{eq:rotation_effect}--\ref{eq:rotation_effect_center} can be
found with respect to the parameters $\alpha_{\mathrm{rot}}$, $\beta$, and $\psi$ if
the viewing angle of the jet, $\theta$, is known. Based on the
analysis of the counter jet feature (see
Sect.~\ref{sc:m87:viewing_angle}), a viewing angle $\theta =
18\degree$ can be adopted for this purpose. In the subsequent
analysis, we will also test independently this hypothesis considering
an equal contribution of the far-side and near-side of the jet in the
central stream velocity ($\psi = 0$).

We use a least-square fitting technique to solve
Eqs.~\ref{eq:rotation_effect}--\ref{eq:rotation_effect_center}. Robust
estimates of the fit uncertainties are obtained from a Monte Carlo
simulation with 10,000 trials using the input parameters
drawn from Gaussian distributions with a mean and standard deviation
provided by the fitted value and the formal fit uncertainty for a given
fitted parameter. The resulting output distribution of the fitted
values also follows a Gaussian distribution, for each parameter of the
fit. We therefore determine the uncertainty of the fitted
parameters by computing the standard deviation of this distribution.

\begin{table}
\centering
\caption{Parameters for jet rotation estimated in region A ($0.5\ \mathrm{mas}
\le z_{\mathrm{obs}} \le 1\
\mathrm{mas}$ ) and region B $1\ \mathrm{mas} \le z_{\mathrm{obs}} \le 3\ \mathrm{mas}$ for $\theta = 18 \degree$.}
\label{tab:m87_rot_parameter_all_18}
\begin{threeparttable}
\begin{tabular}{cccc}
\toprule
Parameter & Region A & Region B, slow & Region B, fast\\
\midrule
$\alpha_{\mathrm{rot}} [\degree]$ & $7.5 \pm 2.3$  & $-9.6 \pm 4.3$ & $4.5 \pm 2.5$\\
$\beta_{p}$    & $0.55 \pm 0.03$ & $0.54 \pm 0.15$ & $0.9245 \pm 0.003$\\
$\beta_{\phi}$ & $0.08 \pm 0.02$   & $-0.10 \pm 0.05$ & $0.07 \pm 0.04$\\
\bottomrule
\end{tabular}
\begin{tablenotes}
  \small
  \item \textbf{Note:} $\alpha_{\mathrm{rot}}$ -- viewing angle modification due to jet
  rotation; \item $\beta_p$, $\beta_{\phi}$ -- poloidal (axial) and toroidal
  (azimuthal) component of the flow speed.
\end{tablenotes}
\end{threeparttable}
\end{table}

\begin{figure}
    \centering
    \includegraphics{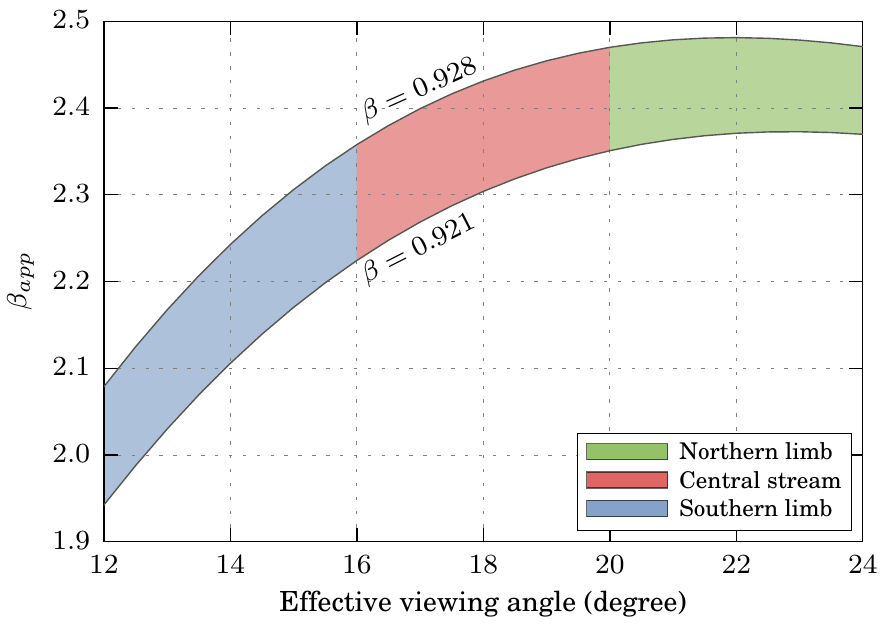}
    \caption{\label{fig:bapp_spread_rotation}
    Effect of jet rotation on the range of apparent velocities observed
    in
    the northern limb (green area), southern limb (blue area) and the central
    stream (red area) assuming the same intrinsic speeds in these three regions.
    With a viewing angle $\theta \sim 18\degree$ and $\alpha_{\mathrm{rot}} \sim 4\degree$, we expect an
    effective
    viewing angle in the range 20 -- 24$\degree$ for the northern limb stream
    and
    in the range 12 -- 16$\degree$ for the southern limb stream.
    }
\end{figure}

The rotation parameters are calculated using the apparent velocities
estimated from the SCC analysis in region A ($0.5\ \mathrm{mas} \le
z_{\mathrm{obs}} \le 1\ \mathrm{mas}$), and independently for the slow
and fast component identified in region B ($1\ \mathrm{mas} \le
z_{\mathrm{obs}} \le 3\ \mathrm{mas}$). The results are summarized in
Table~\ref{tab:m87_rot_parameter_all_18}, assuming a viewing angle
$\theta = 18 \degree$. Evidence for a clockwise rotation of the jet is
found in region A with a $\sim 4 \sigma$ significance, and for the
fast component of region B. In this last case, we verify that $\psi$
is compatible with the inequality of
Eq.~\ref{eq:effective_viewing_angle}: $\psi = -0.2 \pm 5\degree$ while
$\Theta = 2.7\degree$ at $z_{\mathrm{obs}} \sim 2.5$ mas.
Additionally we can obtain an independent estimate of the viewing
angle of the jet $\theta$ assuming $\psi = 0$:
\begin{equation}
  \label{eq:viewing_angle_rotation} 
  \theta = 19.2 \pm 3.7 \degree 
\end{equation}
which supports very well our initial hypothesis. On the contrary, the slow
component of
region B is consistent with a counterclockwise
rotation, which may indicate that this kinematic component reflects
pattern motion rather than true physical
rotation of the flow.

Plasma instability can produce pattern motions with a substantial
azimuthal velocity component which can both co-rotate and
counter-rotate with the flow
rotation~\citep{bodo_on_1996,porth_threedimensional_2013}. This might
be the case for the slow component detected in region B, which has
already been associated with an instability pattern (see
Sect.~\ref{sc:jet_stratification_model}). A counterclockwise rotation
of the instability pattern is also inferred for the jet in M\,87 from
the analysis of the internal structure of the flow on kiloparsec
scales \citep{lobanov_internal_2003}. We adopt therefore this
explanation in our subsequent analysis of the physical conditions in
the jet.

\begin{figure*}
    \centering
    \includegraphics{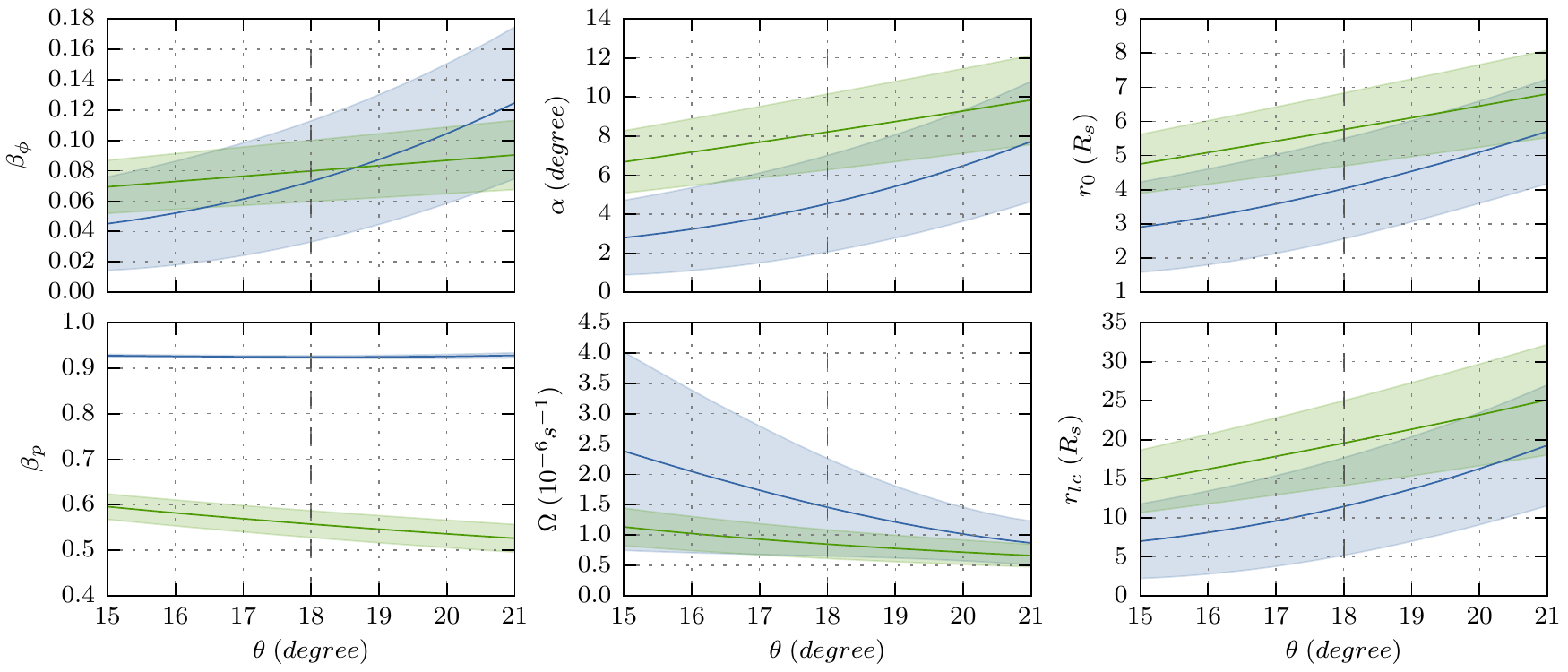}
    \caption[Parameters for jet rotation obtained in regions between 0.5
    and 1 mas and between 1 and 4 mas at different effective viewing angle
    $\theta$.]
    {Parameters for jet rotation obtained in Region A(green)
      and the fast component in Region B (blue) evaluated over a
      range of jet viewing angles. The shaded region corresponds to
      one sigma uncertainty.}
    \label{fig:m87_jet_rotation_all}
\end{figure*}

Jet rotation (irrespectively of its origin) can also explain the different
broadening of the SCC peaks observed in Fig.~\ref{fig:m87_gncc_full_1_4mas} for
the fast velocity component detected in Region B. The observed gradual increase
in the peak width going from the northern limb to the southern limb agrees well
with the hypothesis of a different effective viewing angle in the three parts of
the jet.  For the same intrinsic speed in all three streams of the jet,
Fig.~\ref{fig:bapp_spread_rotation} demonstrates that a smaller effective
viewing angle (southern limb), will result in a larger range of apparent
velocities, while on the contrary, a larger effective viewing angle (northern
limb) will result in a smaller range of apparent velocities.

The smaller effective viewing angle to the flow in the southern limb
relative to the northern limb resulting from rotation would result in
differential Doppler boosting.  Instabilities and the potential slower outer
layer make it difficult to accurately measure jet rotation from the brightness
ratio between the two limbs. However, the mean brightness ratio between the
northern and southern limbs is 0.56, and close to the 0.4 value predicted to
accompany our estimated fast layer jet rotation.

\subsubsection{Interpretation of the jet rotation}

If the azimuthal components of the jet velocity determined above reflects
physical rotation of the jet, they can be related to specific mechanisms of
formation and acceleration of relativistic flows in AGN. In the context of an
MHD jet launched by magneto-centrifugal forces, the initial toroidal velocity is
associated either with the Keplerian speed at the launching location in the
accretion disk~\citep[][BP mechanism]{blandford_hydromagnetic_1982} or with the
spin of the central engine~\citep[][BZ
mechanism]{blandford_electromagnetic_1977}. Conservation of angular momentum
will result in a slow decrease in the toroidal velocity beyond the light
cylinder~\citep{vlahakis_theory_2015}. Observational signature of jet rotation
has been previously reported in jets from young stellar objects (YSO), based on
spectroscopic radial velocity
analysis~\citep[e.g.][]{anderson_locating_2003,choi_rotation_2011}, and has been
successfully associated with an MHD process, giving a constraint on the jet
launching location. Jet rotation has so far not been directly detected in
AGN jets, with the reported rotational speeds clearly related to the rotation of
patterns inside the flow~\citep[e.g.][]{lobanov_internal_2003,agudo_nrao150_2007,
agudo_oj287_2012,cohen_bllac_2016} resulting from plasma instability in the
flow. If our measured rotation is associated with conserved angular momentum in
the jet and if the jet is launched from the accretion disk via the BP mechanism,
its direction should be the same as the rotational direction of the disk itself.
Using spectroscopic HST measurements, \cite{harms_hst_1994} found that the
ionized gas associated with the accretion disk in M\,87 rotates in a clockwise
direction. This is the observed rotation direction found in region A and for the
fast component in region B. It is also interesting to note that the toroidal
velocity measured closer to the central engine is greater than that measured
farther away, which is what the conservation of angular momentum predicts. We
will now try to associate this observed rotation with the MHD properties of the
flow, following~\cite{anderson_locating_2003}.

In the approximation of a cold steady MHD jet, the equations of
conservation of total (magnetic + matter) specific energy $\mu$ and angular
momentum $L$ can be
combined, eliminating the toroidal magnetic field $B_\phi$ and the
distribution of mass flux at the inlet of the flow $\eta$, and
yielding the jet specific momentum $J$:
\begin{equation}
\begin{aligned}
&\mu = \gamma - \frac{r \Omega B_{\phi}}{c \eta} \\
&L = r c \gamma \beta_{\phi} - \frac{c r B_{\phi}}{\eta} \\
&J = \mu c^2 - \Omega L = \gamma c^2 - \Omega r c \gamma \beta_{\phi}
\end{aligned}
\end{equation}
The jet specific momentum, $J$, is conserved along any particular
field line and can be used to obtain an expression for the angular
velocity, which is also a conserved quantity along a given field line:
\begin{equation}
\label{eq:angular_velocity}
\Omega = \frac{c}{r \beta_{\phi}} \left(1 - \frac{J}{c^2 \gamma}\right)
\end{equation}
At the origin of the outflow, we can assume that $\beta_{\phi}$ is well
below the speed of light ($\Omega r_0 \ll c$, $\beta_{\phi, \mathrm{in}} \ll 1$), which
gives $J \simeq \gamma_\mathrm{in} c^2 \simeq c^2$. Knowing the toroidal velocity and
Lorentz factor of the flow, Eq.~\ref{eq:angular_velocity} can be used to
estimate the angular velocity and the associated light cylinder radius 
($r_\mathrm{lc} = c / \Omega$). Assuming Keplerian rotation in the accretion
disk, we can also obtain the launching location of a field line: 
\begin{equation}
r_0 = \left( \frac{G\,M_{BH}}{\Omega^2} \right)^{1/3}\,,
\end{equation}
with $M_{BH}$ denoting the mass of the central engine.

\begin{table}
\centering
\caption{Estimated angular velocity and launching location for Region
A and for
the fast component in Region B, with $\theta = 18 \degree$.}
\label{tab:m87_mhd_rot_parameter_all_18}
\begin{tabular}{ccc}
\toprule
Parameter & Region A & Region B, fast\\
\midrule
$\Omega \,[10^{-6}\ s^{-1}]$ & $0.9 \pm 0.3 $  & $1.5 \pm 0.8$\\
$r_{0} \ [R_\mathrm{s}]$  & $5.7 \pm 1.1$   & $4.0 \pm 1.5$\\
$r_\mathrm{lc}\ [R_\mathrm{s}]$  & $19 \pm 5$      & $11 \pm 6$\\
\bottomrule
\end{tabular}
\end{table}

The resulting parameters obtained for region A and for the fast
component in region B are summarized in
Table~\ref{tab:m87_mhd_rot_parameter_all_18} for a viewing angle of
$18\degree$. For completeness, the respective rotation parameters and
angular velocity estimates for viewing angles between $15\degree$ and
$21\degree$ are also provided in
Fig~\ref{fig:m87_jet_rotation_all}. Continuity of the flow between
regions A and B, which can also be inferred from the VLBA images
(see Fig.~\ref{fig:m87_velocity_field}), indicates that both these
regions can represent the same field line, and hence should yield
similar underlying rotational properties of the flow. Indeed, the
estimates listed in Table~\ref{tab:m87_mhd_rot_parameter_all_18} for
the two jet regions agree within the uncertainties, and a weighted
average of the two measurements gives 
\begin{equation} \begin{aligned}
    \Omega &= 1.1 \pm 0.3 \times 10^{-6} s^{-1}\\ r_0 &= 4.8 \pm 0.8\
    R_\mathrm{s} \,. \end{aligned} \end{equation}

\begin{figure*}
    \centering
    \includegraphics{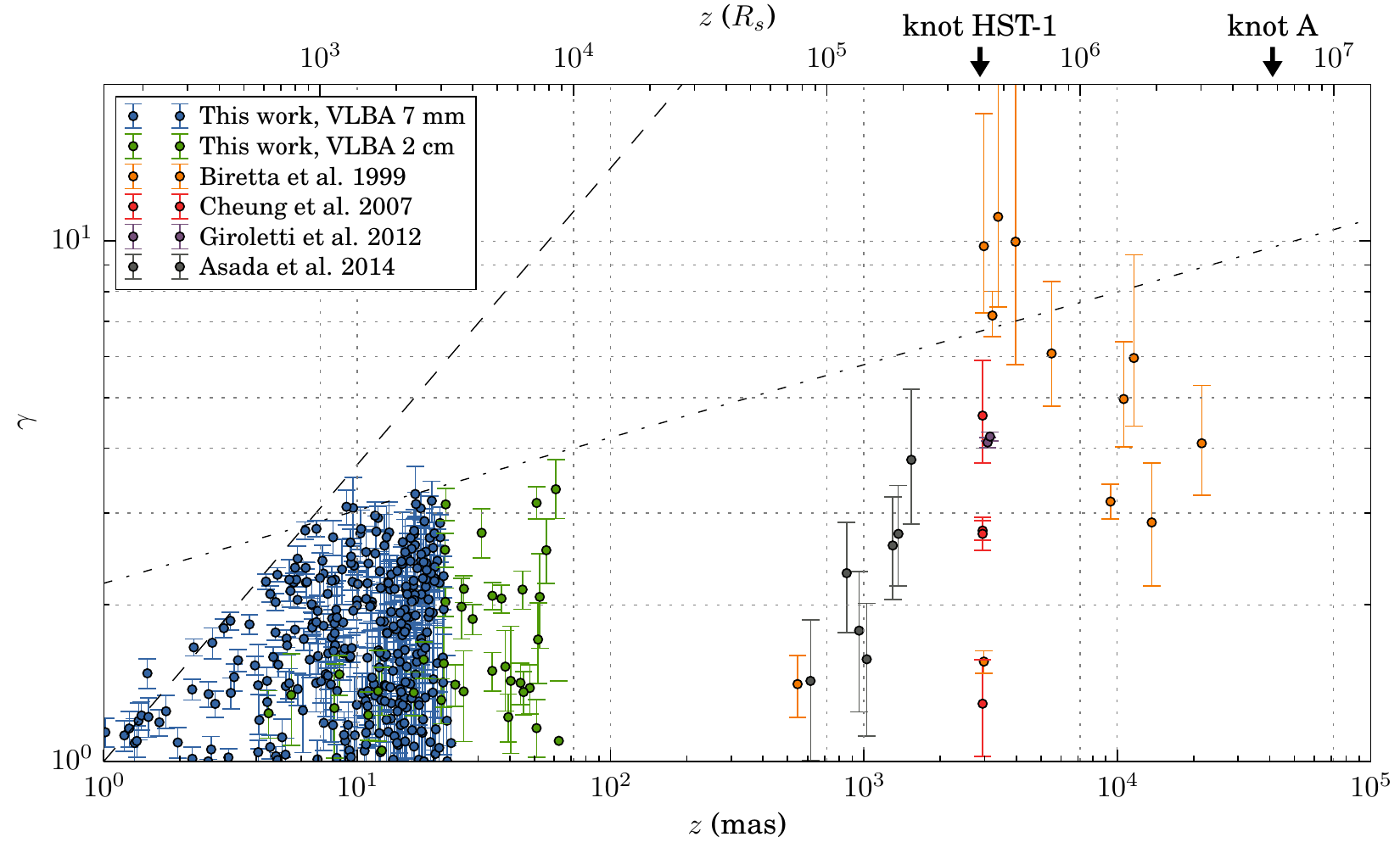}
    \caption[Acceleration profile of the jet in M\,87 from WISE analysis and
    historical measurements.]
    {Acceleration profile of the jet in M\,87. We plot the Lorentz factor
    measured from WISE analysis of VLBA images at 7 mm (blue), 2 cm (green),
    from~\cite{cheung_superluminal_2007} using VLBA at 1.7 GHz (red), 
    from~\cite{giroletti_kinematic_2012} and~\cite{asada_discovery_2014}
    using EVN at 1.7 GHz (magenta and gray respectively),
    and from optical images using HST\citep{biretta_hubble_1999} (orange).
    The Lorentz factor and de-projected z axis are computed assuming $\theta =
    18\degree$ ($z \sim 3.2~z_{obs}$).
    Two regimes are found with a linear acceleration up to $z \sim~ 10^{3} R_\mathrm{s}$,
    followed by a slow acceleration up to HST-1}
    \label{fig:m87_full_lorentz_factor_profile3}
\end{figure*}

Several potential difficulties in invoking jet rotation in the interpretation of
some observed properties in YSO outflows have been discussed in the literature.
Numerical simulations have shown that several disruptive effects, for instance
shocks~\citep{fendt_jet_2011} or plasma instability~\citep{staff_hubble_2015}
can indeed affect the rotational signature from YSO jets. These effects should
be mitigated in our case, as our rotation measurements also include a fast
component velocity which should be related to the inner layers of the jet and
thus less affected by KH instability operating near the jet surface  (however
a body mode of the KH instability or current-driven instability could still have
an impact on the observed jet rotation). We remark also that no re-collimation
shocks are observed in the VLBA images of the M\,87 jet and that the jet
continues to be straight and steady up to Knot A at a de-projected distance of
several kiloparsecs. We finally note that the location of the jet launching
region that we derived from the angular velocity measurement is in the inner
part of the accretion disk close to the innermost stable circular orbit (ISCO)
of the black hole, which agrees well with the predication from numerical
simulations of magnetized relativistic
jets~\citep{meier_magnetohydrodynamic_2001,komissarov_magnetic_2007}.

%

\subsection{Jet collimation and acceleration}
\label{sc:m87:jet_collimation_acc}

Analytical and numerical simulation MHD work has shown that
the processes of jet collimation and acceleration are related and
occur on the same spatial scales in relativistic flows. Most of the
flow acceleration might result from the so-called ``magnetic nozzle''
effect~\citep{li_electromagnetically_1992,vlahakis_theory_2015}. This
effect is at work whenever the flow collimates so that the distance
between streamlines increases faster than the radius of the flow. This
results in a small deviation of the local poloidal magnetic field from
the mean magnetic field and this drives the acceleration. Since this
deviation is small, the fluid is only slowly accelerated by a small
residual force and the acceleration zone can extend over a large
distance. Both numerical simulations and observational evidence
suggest that jet acceleration continues up to distances of
$10^{4}\text{--}10^{5}\ R_\mathrm{s}$ from the central engine~\citep
{vlahakis_magnetic_2004,lobanov_3c345_1996,lobanov_3c345_1999,homan_mojave_2015,lee_3mm_2008,lee_3mm_2016}.

The ideal MHD equations describing plasma jets may be written as a set
of two non-linear differential equations describing the equilibrium of
forces perpendicular (Grad-Shafranov or transfield equation) and
parallel (Bernoulli equation) to the magnetic surface
\citep{camenzind_centrifugally_1986,fendt_ultrarelativistic_2004}. A
complete analytic solution of these equations can be obtained only in
specific, restricted cases, ({\em e.g.}, self-similar solutions). The
force-free approximation is often used to describe jets in the near
zone, inside the light cylinder ($r \lesssim
r_\mathrm{lc}$)~\citep[e.g.][]{blandford_electromagnetic_1977}. At larger
radial separations, different sets of approximations may need to be
used.

In a recent paper, \cite{lyubarsky_asymptotic_2009} obtained
asymptotic solutions for the case of a cold, ideal, Poynting flux
dominated ($\sigma \gg 1$) MHD jet in the far-zone ($r \gg
r_\mathrm{lc}$). In these solutions, different sets of approximations are
used, depending on the decay of the external pressure confining the
jet. In AGN, this confinement can be achieved through a variety of
forces including gas pressure from the ambient medium, ram pressure from
the outer, subrelativistic wind and the stress of a magnetic field
anchored in the disk. In general, evolution of the confining
pressure can be approximated by a power law:
\begin{equation}
P_\mathrm{ext} = P_0 z^{-\kappa}
\end{equation}

\cite{lyubarsky_asymptotic_2009} found two different collimation
and acceleration regimes for the jet, depending on 
specific ranges of the power index $\kappa$. 

If $\kappa < 2$, the jet is in the so-called \textit{equilibrium}
regime.  The residual of the hoop stress and the electric force are
counterbalanced by the pressure of the poloidal magnetic field so that,
at any distance from the source, the structure of the flow is the
same as the structure of an appropriate cylindrical equilibrium
configuration. The jet shape is given by a power law $r \propto
z^{k}$, with $k = \kappa / 4$ converging to a cylindrical shape at
large distance. The acceleration is linear, meaning that it is
proportional to the jet radius: $\gamma \propto r \propto z^ {k}$.

If $\kappa > 2$, the jet is at first in a regime similar to the
equilibrium regime described above. The jet shape is parabolic $r
\propto z^{k}$ and the acceleration is linear. At some distance, the
pressure of the poloidal magnetic field becomes small, and the flow
could be viewed as a composition of coaxial magnetic loops. This
regime is called \textit{non-equilibrium}.  The flow reaches
ultimately a conical shape and the acceleration is determined by the
curvature of the flow, $\gamma \propto z^{(\kappa - 2) / 2}$.

We found in Sect.~\ref{sc:jet_collimation} that the evolution of the
cylindrical cross-section of the jet in M\,87 is described by a power
law with an index $k \sim
0.56\text{--}0.60$. \cite{asada_structure_2012} found that the M\,87
jet reaches a conical shape at kiloparsec scales, similarly to the
$\kappa > 2$ case described above. To determine the acceleration
profile, we compute the Lorentz factors for the velocity field
obtained from WISE analysis at 7~mm
(Sect.~\ref{sc:m87:wise_velocity_7mm}) and 2~cm
(Sect.~\ref{sc:m87:wise_velocity_2cm}), using a viewing angle $\theta
= 18\degree$. We complement the radio measurements with the speeds
determined from optical and radio images of the kiloparsec-scale jet
in M\,87
~\citep{biretta_hubble_1999,cheung_superluminal_2007,giroletti_kinematic_2012,asada_discovery_2014}.
The combined evolution of the Lorentz factor plotted in
Fig.~\ref{fig:m87_full_lorentz_factor_profile3} indicates two
different regimes for the acceleration. The acceleration is at first
linear with $\gamma \propto r \propto z^{0.58}$ up to $z \sim 8$\,mas
($\sim 10^3\,R_\mathrm{s}$), and then it becomes weaker, with $\gamma \propto
z^{0.16}$, agreeing well with the measurements obtained at larger
scales, up to the knot HST-1. The jet appears to enter the
non-equilibrium regime ($\kappa >2$) on these scales. One must however
be cautious about this interpretation. The scaling relations obtained
for the $\kappa > 2$ case are obtained for a conical jet. However the
jet in M\,87 transits to a conical shape only at $z \sim 10^{3}$
mas~\citep{asada_structure_2012}. An alternative explanation for this
slower acceleration can be offered by an early saturation of Poynting
flux conversion. In this case, the conversion should be nearly
complete at $z \sim 10^{3} R_\mathrm{s}$, and acceleration would then continue
slowly until full conversion is achieved.

In the following, the acceleration and collimation of the jet in M\,87
is tested against the magnetically accelerated jet model. Two
different methods are used to solve the MHD equations. The first
method is based on the analytical approximations obtained
by~\cite{lyubarsky_asymptotic_2009} which describes Poynting flux
dominated flows and is mainly used to test the initial acceleration
phase. A second method, that relaxes the assumption of a Poynting
dominated jet, is used to investigate the second acceleration phase,
the saturation of Poynting flux conversion and the transition to a
kinetically dominated jet.

\begin{figure}
    \centering
    \includegraphics{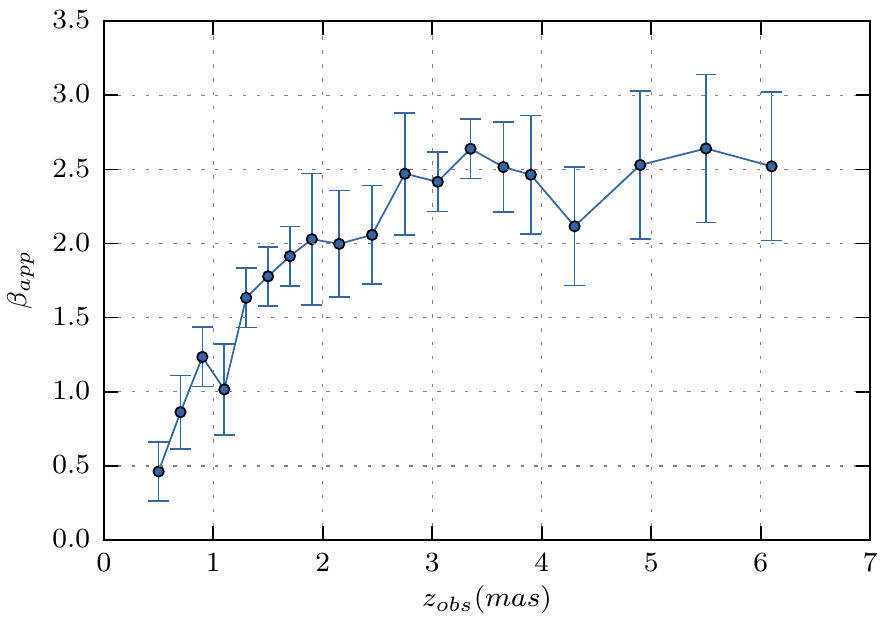}
    \caption{Evolution of the apparent velocity with distance from the core of
    the fast velocity component in the jet sheath.}
    \label{fig:fast_speed_extracted}
\end{figure}

The velocity profile corresponding to the fast velocity component of the flow,
which has been associated with the sheath of the jet in
Sect.~\ref{sc:jet_stratification_model}, is obtained by computing the mean of
the fastest 10\% of the speeds measured within individual bins of 0.1 mas in
size. To improve the robustness of the velocity determination, resampling is
done using bins of 0.2 mas for $0.4 \le z_{\mathrm{obs}} < 2$ mas, 0.3 mas for
$2 \le z_ {\mathrm{obs}} < 4$ mas and finally 0.6 mas for $z_{\mathrm{obs}} \ge
4$ mas. The result is plotted in Fig.~\ref{fig:fast_speed_extracted}. We
ascribe these measured velocities to a single field line (located on a
single magnetic flux surface) with a profile corresponding to the collimation
profile of the sheath as determined in Sect.~\ref{sc:jet_collimation}. Under
this assumption, and assuming $\Omega = \text{const}$ along this field line, we
can conveniently use variables scaled to the light cylinder radius,
$R = \Omega\, r/c$ and $Z = \Omega\, z/c$.

\subsubsection{Asymptotic relations in the far-zone Poynting dominated 
approximation}

In the ideal cold MHD approximation, one can further simplify the solutions by
considering the additional approximation of a Poynting dominated
jet ($\sigma \gg 1$) in the far-zone ($R \gg 1,\ \gamma \gg 1$)~\citep{lyubarsky_asymptotic_2009}.

In the \textit{equilibrium} case and for $\kappa > 2$ one can obtain the
following relations for the jet shape:
\begin{equation}
\begin{aligned}
\label{eq:lyu_collimation}
R =& \left(\frac{3\,Z^{\kappa}}{\epsilon_\mathrm{p}}\right)^{1/4} f\\
f =& \sqrt{\frac{2 - \kappa}{\pi}} \left[ \frac{1}{C_1} \cos^2 w + C_1 \left(C_2
\cos w + \frac{\pi}{2 - \kappa} \sin w \right)^2 \right]^{1/2} \\
w =& \frac{2 \sqrt{\epsilon_\mathrm{p}}}{2 - \kappa} Z^{1 - \kappa/2} - \frac{4 - \pi}{2 -
\kappa} \frac{\pi}{4}\,,
\end{aligned}
\end{equation}
with $\epsilon_\mathrm{p}$ describing the ratio of the plasma pressure to the magnetic
pressure:
\begin{equation}
\epsilon_{\mathrm{p}} = \frac{6 \pi p_0}{B_0^2}\,, 
\end{equation}
where $B_0$ and $p_0$ are the characteristic magnetic field and the external
pressure at the light surface, respectively. 
In the \textit{non-equilibrium}
case, the jet shape is asymptotically conical. 

Evolution of the Lorentz factor can also be obtained, depending on the
acceleration regime:
\begin{equation}
\label{eq:lyu_acceleration}
\gamma = 
\begin{dcases}
R & (\text{equilibrium})\,, \\
\frac{Z^{(\kappa - 2) / 2}}{\sqrt{\beta} \Theta_{\mathrm{app}}^{\mathrm{break}}}  &
(\text{non-equilibrium})\,, \\
\end{dcases}
\end{equation}
where $\Theta_{\mathrm{app}}^\mathrm{break}$ is the local
opening angle at the transition between the two regimes. From
Fig.~\ref{fig:fast_speed_extracted}, we estimate that this transition
occurs in the jet in M\,87 around $z_{\mathrm{obs}}^{\mathrm{break}}
\sim 1.7$ mas, which, in return, corresponds to
$\Theta_{\mathrm{app}}^{\mathrm{break}} \sim 10.5\degree$.

\begin{table}
\centering
\caption{Best fit parameters reproducing the observed acceleration and
collimation using asymptotic relations in the far-zone of the Poynting flux
dominated approximation.}
\label{tab:asymptotic_relations_fit}
\begin{tabular}{lr}
\toprule
Parameter & Best fit\\
\midrule
$z_{\mathrm{obs}}^{\mathrm{break}} \,[\unit{mas}]$ & 1.7 \text{(fixed)} \\
$\Theta_{\mathrm{app}}^{\mathrm{break}} \,[\degree]$ & 10.5 \text{(fixed)}\\
$k$  & $0.588 \pm 0.005$     \\
$\Omega \,[10^{-6}\ \unit{s^{-1}}]$ & $0.86 \pm 0.02 $\\
$\epsilon_{\mathrm{p}}$  & $79 \pm 12$   \\
$\theta \,[\degree]$  & $19 \pm 4$ \\
$C_1$  & $0.10 \pm 0.03$   \\
$C_2$  & $-1 \pm 2$ \\
$a_{\gamma}$  & $0.63 \pm 0.13$ \\
$\phi_{\gamma} \,[\degree]$  & $35 \pm 7$ \\
\bottomrule
\end{tabular}
\end{table}

The Lorentz factor and the collimation profile are fitted
simultaneously using the Levenberg-Marquardt least-squares minimization
routine. Uncertainties in the fitted parameters are determined
from the standard error obtained from the estimated covariance
matrix. The collimation profile is modeled using the relation from
Eq.~\ref{eq:lyu_collimation}. The two phases of the acceleration,
before and after $z_{\mathrm{obs}}^{\mathrm{break}}$, are modeled
using the relations from Eq.~\ref{eq:lyu_acceleration} for the
equilibrium and non-equilibrium case respectively.

Some modifications to the Lorentz factor are required for the fit to
converge.  An additional scaling, $\alpha_\gamma$, was introduced
modifying the Lorentz factor so that: $\gamma^{\star} = a_{\gamma}
\gamma$, and the Lorentz factor was phase shifted along the axial
coordinate of the jet by $w^{\star}_{\gamma} = w + \phi_ {\gamma}$, in
order to be able to account for the fact that the maxima and minima
found in the collimation profile do not correspond exactly to the
maxima and minima found in the Lorentz factor profile. The Lorentz
factor for the non-equilibrium case was obtained
by~\cite{lyubarsky_asymptotic_2009} considering a non-oscillating
conical jet shape. Fig.~\ref {fig:fast_speed_extracted} shows that,
even after $z_{\mathrm{obs}}^{\mathrm {break}}$, the Lorentz factor is
still oscillating. We account for this observation by extending, in our
model, the oscillation to the non-equilibrium case, so that
$\gamma^{\star}_{ne} = a_{\gamma} \gamma_{ne} f$.

All parameters are allowed to vary freely, including the viewing angle $\theta$
and the angular velocity $\Omega$. The best fit obtained under these conditions
has a reduced $\chi^2 = 0.24$. In detail, this corresponds to $\chi_{r}^2 =
0.22$ and $\chi_{\gamma}^2 = 0.34$ for the collimation and Lorentz factor fit
respectively. We recall that $\chi_{r}^2 = 0.55$ was found by modeling the
collimation profile as a simple power law. This indicates an improvement, by a
factor of $\sim 2$, of the quality of the fit provided by the Poynting flux
approximation for the jet collimation profile. We note that the relatively small
values obtained for the reduced chi-squared is most probably a consequence of
our conservative choice for the errors in the jet radius and jet velocity.

\begin{figure}
    \centering
    \includegraphics[width=\columnwidth]{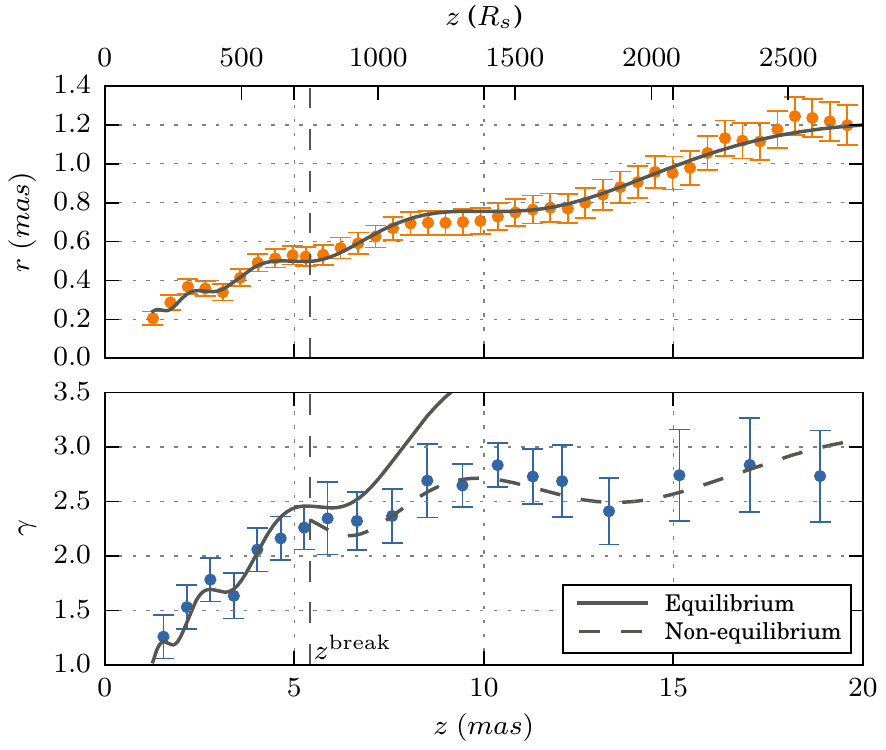}
    \caption[Best fit of the observed acceleration and collimation of the jet in
    M\,87.]
    {Best fit for the observed acceleration (blue) and collimation
      (orange) of the jet in M\,87. The viewing angle and angular
      velocity are independently determined from the fit. Parameters
      of the fitted model are listed in
      Table~\ref{tab:asymptotic_relations_fit}.}
    \label{fig:m87_acceleration_collimation_fit}
\end{figure}

The best fit is plotted in Fig.~\ref{fig:m87_acceleration_collimation_fit}, and
the derived parameters of the model are summarized in
Table~\ref{tab:asymptotic_relations_fit}. It is remarkable that the viewing
angle and the angular velocity agree, within the uncertainties, with the
respective values found from the analysis of the counter jet
(Sect.~\ref{sc:m87:viewing_angle}) and the jet rotation
(Sect.~\ref{sc:m87:jet_rotation}). The oscillations in both the jet width and
the Lorentz factor that we discussed in Sect.~\ref{sc:jet_collimation} are also
particularly well reproduced by the model, and if we extend this solution, we
find that the subsequent Lorentz factor located at $z_{\mathrm {obs}} = 6.8$
mas
and $z_{\mathrm{obs}} = 17.2$ mas, coincide with the two regions of maximum
apparent velocities identified in our WISE analysis of the MOJAVE observations
(Sect.~\ref{sc:m87:wise_velocity_2cm}). \cite{komissarov_stationary_2015}
have
interpreted these oscillations as a standing magneto-sonic wave bouncing across
the jet.

There are several reasons why a phase shift, $\phi_{\gamma}$, and scaling factor
, $\alpha_\gamma$, are required in our modeling. A possible reason for the phase
shift $\phi_{\gamma}$ is that the measured Lorentz factor may come from a
streamline slightly inside the measured jet shape. In this case the maxima in
the Lorentz factor will be offset from the maxima in the jet width, with a shift
proportional to the jet radius~\citep{norman_structure_1982}. The need for the
scaling factor, $\alpha_\gamma$, may be found in the assumption made by
\cite{lyubarsky_asymptotic_2009} of a constant angular velocity for all field
lines. This assumption is usually valid only for a jet launched from the
magnetosphere. For the case of disk launching, we expect differential rotation
and~\cite{tchekhovskoy_simulations_2008} found $\gamma < R$ under these
conditions. We recall here also that relations were derived using
assumptions, specifically $R \gg 1$ and $\gamma \gg 1$, which are only partially
valid in our case.

While the Lorentz factor of the second acceleration
regime is well modeled by the non-equilibrium case, it is not
associated with transition to the conical jet shape that one
should expect theoretically. The apparent opening angle at the
location where the acceleration regime changes is
$\Theta_{\mathrm{app}}^{\mathrm{break}} \sim 10.5\degree$, and therefore
additional collimation is necessary to achieve the final apparent
opening angle $\Theta_{\mathrm{app}}^{\mathrm{kpc}} \sim 3.2\degree$
observed at kiloparsec scales. Early saturation of the Poynting flux
could also result in a quenched acceleration, and would break the
assumptions of a Poynting dominated flux used in this modeling. We will
now investigate this possibility by using another method of solving
the MHD equations.

\subsubsection{Wind solutions}
\label{sc:wind_solution}

Solving simultaneously the full transfield and the Bernoulli equations
is only possible in a restrictive number of cases. The transfield
equation will determine the shape of the field lines while the
Bernoulli equation governs the acceleration of the flow. It is however
possible to solve only the Bernoulli equation by assuming a specific
jet shape~\citep[e.g.][]
{camenzind_centrifugally_1986,fendt_ultrarelativistic_2004,toma_efficient_2013}.
The resulting solution is called a wind solution. It is not strictly
an exact MHD solution, and one should be careful not to over-interpret
it. However it can be used in some specific cases, for example, for
investigating the acceleration efficiency, or in our case, the
acceleration profile.

In the context of a cold ideal MHD jet, the poloidal and toroidal velocities
can be written as follows:
\begin{equation}
\begin{aligned}
\label{eq:beta_p_phi}
\beta_{p} &= \frac{R^2 B_{p}(R)}{\eta (\mu - \gamma)} \left(1 - \frac{1}{R^2}
\left(1- \frac{\gamma_{in}}{\gamma}\right)\right) \\
\beta_{\phi} &= \frac{1}{R} \left(1 - \frac{\gamma_{in}}{\gamma}\right)\,,
\end{aligned}
\end{equation}
where $B_p$ is the poloidal magnetic field, $\mu$ is the total (magnetic +
matter) specific energy, $\gamma_{in}$ and $\eta$ are the Lorentz factor and the
distribution of mass flux at the inlet of the flow, respectively. The quantity
$\eta$ is a field line constant and it is given by the following equation:
\begin{equation}
\eta = \frac{4 \pi \gamma \rho_0 \beta_p c^2}{B_p}\,,
\end{equation}
with $\rho_0$ describing the rest-mass density. We also define the ratio of the Poynting to the
matter energy flux, also called magnetisation parameter, as:
\begin{equation}
\sigma = \frac{\mu - \gamma}{\gamma}
\end{equation}
The Bernoulli equation is obtained from:
\begin{equation}
\beta_{p}^2 + \beta_{\phi}^2 + \frac{1}{\gamma^2} = 1\,.
\end{equation}
Substituting $\beta_{p}$ and $\beta_{\phi}$ using Eq.~\ref{eq:beta_p_phi}, one
can reduce the Bernoulli equation to a quartic equation for $\gamma$. To
solve it, one has to prescribe a model for the poloidal magnetic field $B_p$,
which turns into a model for the flux function.

\cite{toma_efficient_2013} recently derived a set of models for $B_p$ based on a
realistic flux function $\Psi(r, z)$ which describes parabolic field lines appropriate
for M\,87:
\begin{equation}
\label{eq:flux_function}
Z + \zeta(\psi) = A(\psi) (R - \varpi(\psi))^{a(\psi)}\,,
\end{equation}
where $\psi = \Psi / \Psi_0$ and $\Psi_0$ is the total magnetic flux, and 
$\zeta(\psi)$, $A(\psi)$, $\varpi(\psi)$, and $a(\psi$) are arbitrary functions  of $\psi$ \citep{toma_efficient_2013}. The
poloidal magnetic flux is then obtained recalling that
$B_p = (\nabla \Psi \times \phi) / r$, and this yields
\begin{equation}
\begin{aligned}
B_p(R) =& \frac{\Psi_0}{R^2} \left( \frac{\zeta'}{Z + \zeta} - \frac{A'}{A} - a'
\ln(R - \varpi) + \frac{a \varpi'}{R - \varpi} \right)^{-1} \\
&\frac{R}{R - \varpi}
\sqrt{\left( \frac{R - \varpi}{Z - \zeta} \right)^2 + a^2}\,,
\end{aligned}
\end{equation}
with primes denoting derivatives of the respective functions over $\psi$.

At $r \lesssim r_\mathrm{lc}$, the true shape of field lines will probably deviate
from the shape derived from Eq.~\ref{eq:flux_function} where field lines
 are anchored into
the accretion disk or the magnetosphere. However, the approximation should be
valid in the region we are investigating.

In our case of interest, associating the jet sheath with the outer poloidal
field line ($\psi = 1$) and using $\theta = 18 \degree$ and $\Omega = 10^ {-6}$,
we have $\zeta = \varpi = 0$, $A = 3.7$ and $a = 1 / k = 1.73$, corresponding to
the parabolic shape obtained by~\cite{asada_structure_2012}. We also adopt
$\gamma_{in} \rightarrow 1$.


With the assumptions specified above, the Bernoulli equation still contains 7
unknowns: the total specific energy $\mu$, the distribution of mass flux at the
inlet $\eta$, the total magnetic flux $\Psi_0$, and four parameters related to
the geometry of the poloidal field lines ($a'$, $A'$, $\zeta'$ and $\varpi'$).
We can however isolate some parameters and treat different cases separately. We
define three models corresponding to different configurations of the flux
function:

\begin{figure}
    \centering
    \includegraphics[width=\columnwidth]{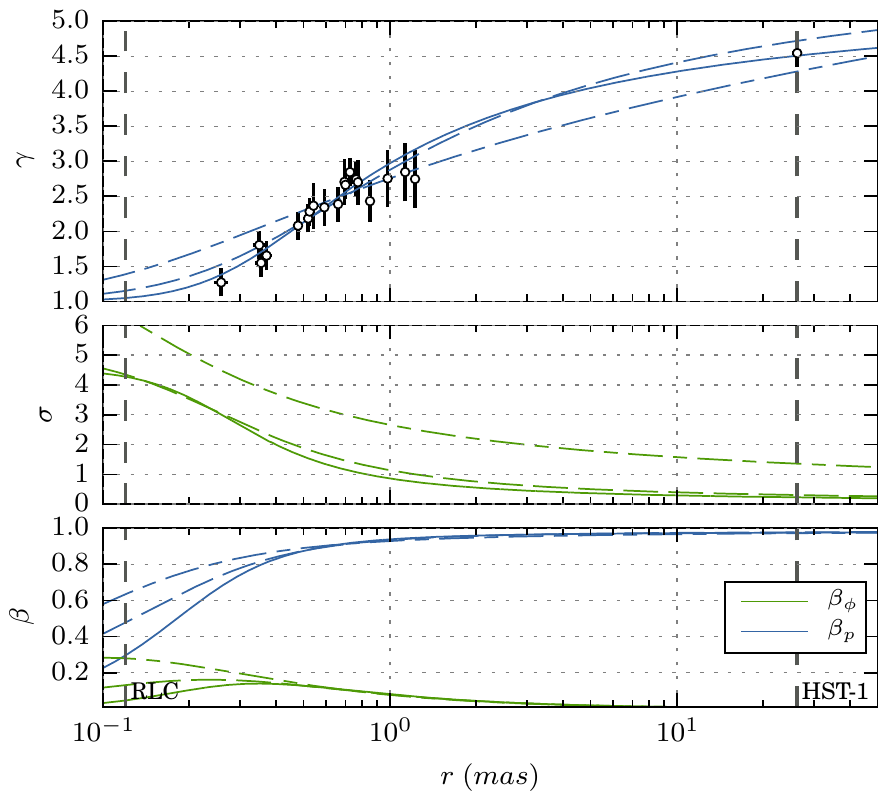}
    \caption[Wind solution for different models of the flux function that reproduces the best the
    observed acceleration.]
    {Wind solution for model 1 (solid line), model 2 (dashed line) and
      model 3 (long dash short dash line) for the flux function that
      best reproduces
      the observed acceleration (black points). The upper and
      middle panels show the total energy (blue) and magnetisation
      parameter (green). The lower panel shows the toroidal (green)
      and poloidal (blue) velocity.}
    \label{fig:bernoulli_fitting_all_velocities_theta18}
\end{figure}

\begin{itemize}[leftmargin=.7in]
  \item [\textbf{Model 1}:] $a' \neq 0$, $A' = 0$ ,$\zeta' \neq  0$, $\varpi' = 0$,
  \item [\textbf{Model 2}:] $a' \neq 0$, $A' = 0$ ,$\zeta' =  0$, $\varpi' \neq 0$,
  \item [\textbf{Model 3}:] $a' \neq 0$, $A' \neq 0$ ,$\zeta' =  0$, $\varpi' = 0$.
\end{itemize}

The Bernoulli equation has two singularities at the Alfv\'enic point
(AP) and at the fast magnetosonic point (FMP). The wind solution in
which $\gamma$ is growing toward infinity has to pass through both
the AP and the FMP.  The Bernoulli equation, being
quartic for $\gamma$, admits four roots. To be valid, the wind
solution needs to pass smoothly from one root to the other crossing
the singularities. This provides an additional constraint on the
solution, and for a given $B_p (R)$ and $\eta$, a unique value of
$\mu$ corresponds to a valid wind solution.  Further examining the
Bernoulli equation, we find that $\Psi_0$ and $\eta$ can not be
separated, and their ratio must therefore be treated as a single
variable. Hence we end up with only three free variables for the three
models described above.

A Nelder-Mead least-squares minimization routine is used to fit the
Bernoulli equation to the measured Lorentz factor evolution. Within
each iteration of the fitting algorithm, $\mu$ is determined so as to
maintain a valid wind solution.

Satisfactory fits were obtained for Models 1 and 2, however we could not obtain
a reasonable fit using Model 3. In order still to be able to analyze the impact
of the parameter $A'$ on the solution, we assigned a small non-zero value to
$\varpi'$ in this model ($\varpi' = 0.05$). The results for the three models are
summarized in Table~\ref{tab:m87_bernoulli_fitting} and corresponding wind
solutions are plotted in
Fig.~\ref{fig:bernoulli_fitting_all_velocities_theta18}.

Fig.~\ref{fig:m87_full_lorentz_factor_profile3} suggests that the
velocities obtained for knot HST-1 and from the 43\,GHz VLBA images
originate from the same field line. To test this hypothesis we also
include to the fit the speed measured at HST-1 by
\cite{giroletti_kinematic_2012}, assuming $r_ {\textrm{HST-1}} =26$
mas~\citep{asada_structure_2012}. This addition does not affect the
reduced $\chi_2$ of the fit, thus supporting our hypothesis.

\begin{table}
\centering
\caption{Parameters for the wind solution that reproduce the observed
acceleration for the three flux function models.}
\label{tab:m87_bernoulli_fitting}
\begin{tabular}{lrrr}
\toprule
Parameter & Model 1 & Model 2 & Model 3\\
\midrule
$\mu$ & 5.5 & 6.2 & 10.1 \\
$L\ (\times 10^{23})\ [\mathrm{SI}]$ & 4.1 & 4.7 & 8.2  \\
$\mathrm{FMP}\ [R_\mathrm{lc}]$ & 3.23 & 3.23 & 3.65\\
$\mathrm{AP}\ [R_\mathrm{lc}]$  & 0.92 & 0.92 & 0.95  \\
$B_0 / \eta\ [\mathrm{SI}]$ & 1.5 & 2.9 & 7.8  \\
\midrule
$\chi^2$ & 0.41 & 0.62 & 1.6 \\
\bottomrule
\end{tabular}
\begin{tablenotes}
  \small
  \item \textbf{Note:} $B_0 = B_p(r_\mathrm{lc})$ -- characteristic magnetic field at
  the light cylinder;
\end{tablenotes}

\end{table}

Models 1 and 2 best reproduce the measured acceleration. Jet parameters derived
with these models are also similar. In Model 3, a higher value of $\mu$ was
required, and the resulting conversion from Poynting flux to kinetic energy is
slower. In all three cases, the Alfv\'enic point is found to be located close to
the light cylinder, in agreement with exact self-similar
solutions~\citep{vlahakis_magnetic_2004}. The fast magnetosonic point is located
at $r \sim 0.4$ mas, corresponding to $z \sim 1$ mas, and in agreement with the
expectation that $\gamma_{\mathrm{FMP}} \simeq \mu^{1/3}$.

A true solution of the Bernoulli equation for the jet in M\,87 would
most likely be represented by a combination of the three models. We
also do not know the exact shape of the field line close to the
central engine, and so we cannot reconstruct the initial acceleration
which might have helped discriminate between the models, as suggested by the
divergence of the poloidal and toroidal velocities observed in
Fig.~\ref{fig:bernoulli_fitting_all_velocities_theta18}.  Our result
however strongly favors a moderate value of $\mu$ with an efficient
transformation of Poynting flux to kinetic energy (Models 1 \& 2).

\subsubsection{Transition to a kinetically dominated jet}
\label{sc:magnetization}

The wind solution obtained for the sheath suggests a total specific
energy at the base of the jet $\mu \sim 6$--$10$. In the first
two models described in Sect.~\ref{sc:wind_solution}, conversion
between magnetic and kinetic energy is efficient, and the equipartition
condition $\sigma \sim 1$ is achieved at:
\begin{equation}
r_\mathrm{eq} \sim 1.5\ \text{mas} \sim 10\ r_\mathrm{lc} \sim 200\ R_\mathrm{s}\,,
\end{equation}
corresponding to
\begin{equation}
z_\mathrm{eq} \sim 20\ \text{mas} \sim 2800\ R_\mathrm{s} \sim 1.6\ \mathrm{pc}\,.
\end{equation}

Results from MHD simulations of AGN jets \citep{komissarov_magnetic_2007}
indicate $r_\mathrm{eq} \sim 30\ r_\mathrm{lc}$.  Considering the choice of
total energy $\mu \sim 18$ used for these simulations, our results for Model 1
and Model 2 are qualitatively similar to the results
of~\cite{komissarov_magnetic_2007}.

The conversion of Poynting flux to kinetic energy is less efficient
for our Model 3, and equipartition is only achieved at $r_\mathrm{eq}
\sim 40\ \text{mas} \sim 300\ r_\mathrm{lc}$. However this model is altogether
less favorable as it does not reproduce the observed acceleration as
well as the first two models.

\subsection{Nature of the Spine}
\label{sc:nature_spine}

In Sect.~\ref{sc:jet_stratification_model}, we identified two
primary velocity components in the jet. Further analysis has shown
that the slow component reflects either pattern speed of a plasma
instability or the bulk speed of an outer slower wind. The fast
velocity component has been successfully associated with the bulk
speed along a magnetic field line connected to the observed jet
shape, and we have obtained a wind solution for it. An inner
deboosted part of the jet could therefore be associated with an even
faster speed along a field line located closer to the jet axis.

\begin{figure}
    \centering
    \includegraphics{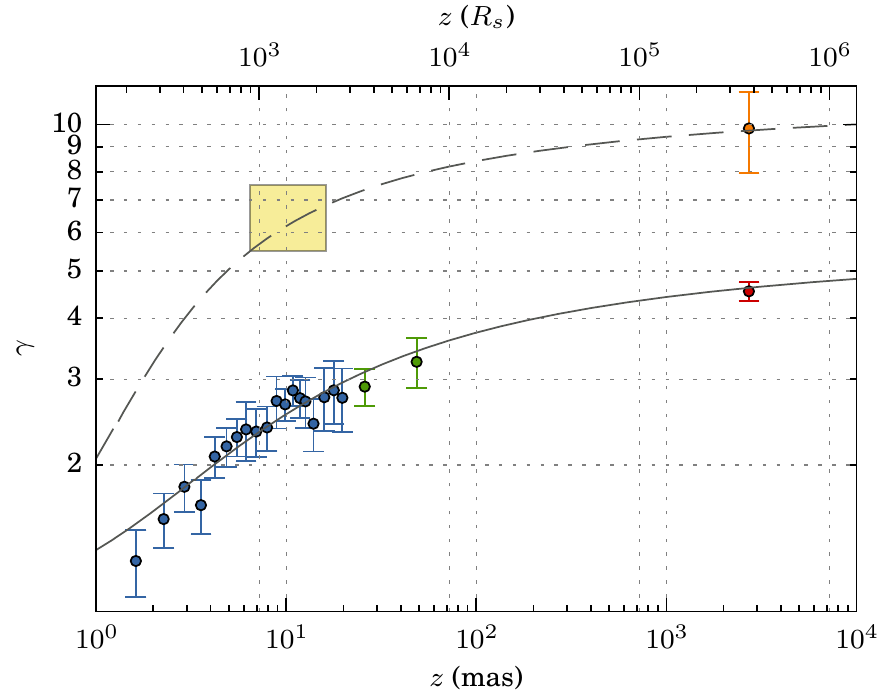}
    \caption[Wind solution for the spine and the sheath.]
    {Wind solution for the spine (dashed line) and the sheath (solid
      line). The solution for the spine fits well the inferred
      velocity of the deboosted structure observed in the 43\,GHz VLBA
      maps (yellow box) and the velocity observed at HST-1 in the
      optical band (orange point). The solution for the sheath fits
      the fast velocities measured in the 43\,GHz (blue points) and
      15\,GHz (green points) VLBA data, and at HST-1 in the radio band
      (red point). The 15\,GHz VLBA data points are binned averages
      of the fastest observed apparent velocities from our WISE
      analysis.}
    \label{fig:wind_sol_sheath_spine_thet18}
\end{figure}

The limb brightened structure found in the M\,87 jet essentially at all scales
might be explained by the combination of an enhanced magnetic field at the jet
edge, a faster speed close to the jet axis and a lower emissivity in the jet
spine. At kiloparsec scales, kinematic analysis systematically found a higher
apparent velocity in the optical
domain~\citep{biretta_hubble_1999,meyer_optical_2013} compared to the radio
domain~\citep{biretta_detection_1995,cheung_superluminal_2007,giroletti_kinematic_2012}
\footnote{\cite{giroletti_kinematic_2012} also discussed a potential faster
spine component, and later observations of this region by~\cite{hada_evn_2014}
constrained the estimated speed of this component to $\sim 5\,c$.},
indicating that they trace different regions in the jet, and that the inner part
(spine) of the jet is dominant at optical wavelength, as was also suggested
earlier by~\cite{perlman_optical_1999}.

In the stacked-epoch image of the jet in M\,87, we measure an
intensity ratio between the inner part (the spine) and the edge (the
sheath) $\mathcal{R} \sim 0.5$ at $z_{\mathrm{obs}} \sim 3$ mas. At
this distance, we also observe $\gamma_ {\mathrm {sheath}} \sim
2.7$. Assuming that the lower intensity near the jet axis is mainly
due to deboosting of the spine, we can calculate an estimate of
the Lorentz factor of the spine necessary to reproduce the observed
intensity ratio using:
\begin{equation} 
\mathcal{R} = \frac{I_{\mathrm{spine}}}{I_{\mathrm{sheath}}} = \left[\frac{\gamma_
{\mathrm {sheath}} (1 - \beta_{\mathrm{sheath}}
\cos(\theta))}{\gamma_{\mathrm{spine}} (1 - \beta_{\mathrm{spine}} \cos
(\theta))}\right]^{2 - \alpha}
\end{equation} 
Solving this equation, we obtain $\gamma_{\mathrm{spine}} \sim
6.5$. If we combine this estimate with the speed observed at HST-1 in
the optical and assume that they are connected to a single field line,
we can find a wind solution that would satisfy these observations. To
obtain an accurate estimate of the velocity at HST-1 in the optical,
we take the weighted mean of the 4 apparent velocities measured by
\cite {biretta_hubble_1999} at HST-1, which gives us
$\gamma_\mathrm{spine} = 9.8 \pm 1.9$.

Starting with the best solution found in Sect.~\ref{sc:wind_solution}
using Model 1, we set $\Omega = 2.5 \times 10^{-6}\ \unit{s^{-1}}$
corresponding to a field line anchored to a launching location closer
to the ISCO ($r_0 \sim 2.8\ R_\mathrm{s}$), $\mu = 12$, and $A = 6.6$ so that
$r_{\mathrm{spine}}(z) = 0.5 r_{\mathrm{sheath}} (z)$. We then look
for a valid wind solution that would fit the two estimates of the
Lorentz factor of the jet spine. The result, plotted in
Fig~\ref{fig:wind_sol_sheath_spine_thet18}, indicates that
while the sheath field line can be associated with the velocity
measured in the radio at HST-1, the spine field line connects well the
inferred velocity of the deboosted structure in 43\,GHz VLBA maps and
the velocity measured at HST-1 in the optical band.

While this model would explain the limb brightened structure at
$z_{\mathrm{obs}} \gtrsim 2\ \mathrm{mas}$, the decrease in spine speed
close to the core would result in an accompanying increase in Doppler boosting
of the spine.  In the simplest interpretation a ridge brightened structure at
$z_{\mathrm{obs}} \sim 1\ \mathrm{mas}$ with brightness ratio $\mathcal{R}
\sim
5$ would occur.  This is not observed in the 7 mm VLBA observations, or in the 3
mm Global VLBI Array (GMVA) maps~\citep{hada_3mm_2016}. This lack of ridge
brightening suggest a low spine emissivity as the reason for the observed limb
brightening down to the smallest core separations.

\subsection{Launching mechanism}
\label{sc:launching_mechanism}

The angular velocity of a sheath field line has been obtained 
independently from the observed flow rotation and from the modeling of
jet acceleration and collimation. The two estimates agree remarkably
well, within the uncertainty, with value $\Omega \sim 10^{-6}\
\unit{s^{-1}}$. This parameter can be used to discriminate between the
two major mechanisms that operate during jet launching. In the
Blandford-Payne model, the jet is launched from the accretion
disk and the field line angular velocity depends on the distance from
the black hole at which this field line is anchored. If we assume a
Keplerian rotation in the accretion disk, this angular velocity corresponds to
$r_0 \sim 5\
R_\mathrm{s}$. In the Blandford-Znajek mechanism, the angular velocity
depends on the spin of the central black hole
$a_\mathrm{H}$. Following~\cite{tchekhovskoy_launching_2015}, the black hole
angular speed is:
\begin{equation}
\begin{aligned}
\Omega_H &= \frac{a_\mathrm{H} c}{2 r_\mathrm{H}} \\
r_\mathrm{H} &= \frac{R_\mathrm{s}}{2} (1 + \sqrt{1 - a_\mathrm{H}^2})
\end{aligned}
\end{equation}
One can then obtain the angular speed of the respective field line,
\begin{equation}
\Omega_\mathrm{F} \simeq 0.5 \Omega_\mathrm{H}
\end{equation}

Using 230\,GHz VLBI observations, \cite{doeleman_jetlaunching_2012}
estimated the spin of the black hole in M\,87 to be $a_\mathrm{H} \sim 0.6$
based on a measure of the size of the smallest resolvable structure,
identified as the ISCO. This value corresponds to $\Omega_\mathrm{F} = 2.75
\times 10^{-6}\ \unit{s^{-1}}$ and is a factor 3 times larger than the
angular velocity that we determined. The BP mechanism would therefore
better account for our result, but a BZ launching cannot be firmly
discarded owing to the uncertainty of this measurement. Nevertheless,
the wind solutions obtained for the sheath and the spine imply
$\mu_{\mathrm{spine}} \gtrsim \mu_{\mathrm{sheath}}$, which is a
characteristic of differential rotation as found both analytically
and from numerical
simulations~\citep{lyubarsky_asymptotic_2009,komissarov_magnetic_2007}. Thus
we favor a disk launching at least for the sheath.

\subsection{Mass-loss rate estimate}

The mass-loss rate corresponds to the amount of matter extracted from
the accretion disk and loaded into the jet. It can be obtained from:
\begin{equation}
\dot{M} = \frac{\eta \Psi_0}{4 c}
\end{equation}
The wind solution describing the acceleration of the sheath requires
$B_0 / \eta \sim 2$--$10$, with $B_0$ describing the characteristic
magnetic field at the light cylinder. The mass loss rate can then be
estimated recalling that $\Psi_0 \sim B_0 r_\mathrm{lc}^2$:
\begin{equation}
\dot{M} \sim \frac{\eta B_0 r_\mathrm{lc}^2}{4 c}
\end{equation}
Using 230\,GHz VLBI observations, \cite{kino_magnetization_2015} estimated
the magnetic field at the base of the flow to be about 100 Gauss. Using this
value for $B_0$, we get $\dot{M} \sim 10^{-7}$--$10^{-8}\ M_{\odot}\unit{yr^{-1}}$.

The mass accreted onto the black hole of M\,87 is still undetermined. Using
Chandra X-ray observations, a Bondi accretion rate of $\dot{M}_\mathrm{B} =
0.12\ M_{\odot}\unit{yr^{-1}}$ was estimated by~\cite{matteo_accretion_2003a}.
However, the measured X-ray luminosity, $L_X \sim 7 \times 10^{40} \unit{erg
s^{-1}}$, suggests a mass-loss rate of $\dot{M}_\mathrm{acc} \ll \dot{M}_\mathrm
{B}$.
\cite{kuo_measuring_2014} found an upper limit of $\dot{M}_\mathrm{acc} < 1
\times 10^{-3}\ M_{\odot}\unit{yr^{-1}}$ using Faraday rotation. This
limit would yield $\dot{M} \sim 10^{-4}$--$10^{-5} \dot{M}_\mathrm{acc}$.

Thus, the mass-loss rate is much lower than the total mass accreted
onto the black hole, implying that only a small fraction of the
accreted matter is transported into the jet.

\section{Summary}

In this paper, we have presented a detailed analysis of the
two-dimensional kinematic evolution of the innermost part of the jet in
M\,87. This source has allowed us to investigate jet
formation and propagation on scales of $10^{2}$--$10^{3}
R_\mathrm{s}$. A WISE analysis
\citep{mertens_waveletbased_2015,mertens_detection_2016} was performed
on 43\,GHz VLBA maps observed as part of the M\,87 VLBA movie
project~\citep{walker_vlba_2008,walker_m87_2016}.  We summarize here the main findings
of this analysis:

\begin{enumerate}
\setlength{\itemsep}{0.4em}
\item We have obtained the first complete velocity field of an AGN jet
  at sub-parsec scale, revealing a structured and highly stratified
  jet, with a significant transverse velocity component.

\item The structure of the flow was also investigated using SWD
  decomposition. The shape of the streamline is parabolic, well
  described by a power law $r \propto z^{0.6} $.

\item We observe an oscillation in flow expansion with a spatial
  period increasing with the distance as $\sim z_{\mathrm{obs}}$,
  which is correlated with a similar oscillating pattern in the apparent speed.

\item The viewing angle $\theta \simeq 18\degree$ was obtained using
  the speed of a moving component discovered in the counter jet, and
  the jet to counter jet intensity ratio. This viewing angle was also
  confirmed by the analysis of jet rotation and by the modeling
  of jet acceleration.

\item Flow stratification was analyzed using the SCC method
  \citep{mertens_detection_2016}, revealing a slow, mildly
  relativistic layer ($\beta \sim 0.5 c$) associated either with an
  outer wind or with instability pattern speed, and a fast
  accelerating streamline ($\gamma \sim 2.5$ at $z_{\mathrm{obs}} \sim
  3$ mas)

\item Acceleration and collimation of the flow was modeled as an MHD jet, using
two different methods. We found that   the acceleration is at first linear
$\gamma \propto r \propto   z^{0.58}$, and then saturates, implying a
total specific energy $\mu \sim 6$, and that equipartition
between Poynting flux and kinetic energy is reached at a distance $z_\mathrm{eq}
\sim   3000\ R_\mathrm{s}$, in excellent agreement with previous analytic   and
numerical simulation work. This suggests that the jet is already   kinetically
dominated at a distance of a few parsecs from the central   engine.


\item We could not detect any features that could unequivocally be related to an
inner spine.  We estimated a spine Lorentz factor $\gamma_{spine} \sim 6$ that
would deboost the spine emission sufficiently to explain the intensity ratio
between sheath and spine.  Subsequently we found a wind solution that would
connect this Lorentz factor with the velocity observed at knot HST-1 in the
optical regime. However, this solution breaks down close to the core where
it would predict a ridge brightened instead of a limb brightened jet.  This
suggests a lower intrinsic emissivity for the jet spine.

\item Indication of jet rotation was found from the difference in
  velocities observed in the northern and southern limbs. This
  observation was used to obtain an angular velocity for the field
  line, $\Omega \sim 10^{-6}\ \unit{s^{-1}}$ which, in turn,
  suggests that this streamline is most likely launched in the inner
  region of the accretion disk, at a distance $r_0 \sim 5\
  R_\mathrm{s}$ from the central engine.  However, a BZ scenario in
  which the jet is launched from the magnetosphere of the central engine
  could not be excluded.

\item Finally, an estimate of the mass-loss $\dot{M} \sim
  10^{-7}$--$10^{-8}\ M_{\odot}\unit{yr^{-1}}$ was obtained for
  the sheath, from the wind solution. This corresponds to $\dot{M}
  \sim 10^{-4}$--$10^{-5} \dot{M}_\mathrm{acc}$.

\end{enumerate}

Taken together, the WISE analysis of the jet in M\,87 draws a coherent
picture of a magnetically launched, accelerated, and collimated jet.
Continuation and expansion of the work presented in this paper will
enable testing the results obtained for the jet in M\,87 and
addressing other important aspects of the physics of relativistic
flows. The methods that we have used to model the acceleration and
collimation zone in the M\,87 jet can be improved further to provide
a more detailed physical description of the flow. One possibility
would be to use a more realistic flux function obtained from numerical
simulations. Another possibility would be to attempt to reproduce our
results by performing full 2D or 3D MHD simulations using the physical
parameters estimated for the jet in M\,87. This would allow us to
investigate the inner part of the jet, providing further clues about
the reason for the edge brightened morphology.

The milliarcsecond-scale jet in M\,87 has been observed at 15 GHz as part of the
MOJAVE project. We demonstrated in this work that the fast stream line can be
robustly detected at this frequency with VLBA observations made roughly every
two months.  Obtaining and analysing a well-sampled 15 GHz VLBA dataset on M\,87
would enable us to extend the analysis of the acceleration profile up to $z \sim
10^{4}\ R_\mathrm{s}$. The high resolution provided by the GMVA observations at
86\,GHz could also be critical in confirming the results we obtained. The
structure of the flow could be probed with GMVA at distances as
small as $\sim 10 R_\mathrm{s}$, providing an excellent probe of the
innermost part of the jet.

\begin{acknowledgements} 

This research has made use of data from the MOJAVE database that is maintained
by the MOJAVE team~\citep{lister_mojave_2009}. F.M. was supported for this
research through a stipend from the International Max Planck Research School
(IMPRS) for Astronomy and Astrophysics at the Universities of Bonn and Cologne.
We sincerely thank Y. Kovalev for providing the VLBA images at 2 cm. F.M. thanks
Y. E. Lyubarsky, C. Fendt, K. Toma and A. Tchekhovskoy for helpful discussions.
We would also like to thanks the anonymous referee for the careful review that
improved the quality of this paper. The National Radio Astronomy Observatory is
a facility of the National Science Foundation operated under cooperative
agreement by Associated Universities, Inc.

\end{acknowledgements}

\bibliographystyle{aa}
\bibliography{biblio}

\begin{appendix}

\section{Intermediate-scales wavelet decomposition}
\label{sect:iwd}

Canonical wavelet transformations are performed on a set of scales
ranging as powers of two of the initial, smallest scale used.  The
ability of the multi-scale cross correlation
\citep[MCC;][]{mertens_waveletbased_2015} procedure to cross identify
significant structural patterns (SSPs) in pairs of images relies on a
smooth transition between scale $j$ and scale $j+1$ of the wavelet
transform employed.  This condition sets limits on detecting different
speeds at different scales, and hence providing an accurate assessment
of flow stratification traced by overlapping optically thin features
with different intrinsic speeds \citep{mertens_detection_2016}. For
any two adjacent scales of the wavelet transform, this limit is on the
order of $2^{j}$~\citep{mertens_waveletbased_2015}. As the
  velocity difference between two overlapping features becomes
  comparable or smaller than this limit, matching becomes more
difficult. To improve cross identification of the individual features,
one can construct a finer scaling function for which the wavelet
decomposition will cover intermediate scales. This is performed by
introducing an intermediate-scale wavelet decomposition (IWD). In a
canonical wavelet transform, the coefficients $w_j$ at a scale $j$ of
the wavelet decomposition contains information on spacial scales
between $2^{j-1}$ and $2^{j}$. In the IWD, the coefficients at a given
scale $j$ contain information on spacial scales between
$1.5\times2^{j-1}$ and $1.5\times2^{j}$. The MCC is then performed by
inserting the scale $j$ of the IWD between the scales $j-1$ and scale
$j$ of the SWD. 

The IWD is realized in the WISE analysis by introducing an appropriate
scaling function.
The \textit{B-spline} scaling function used in WISE has the following
discrete low pass filter associated with it:
\begin{equation}
\begin{pmatrix} 
\slfrac{1}{4} & \slfrac{1}{2} & \slfrac{1}{4}\,.
\end{pmatrix}
\end{equation}
For the IWD, a modified \textit{B-slpine 3} scaling function is used
which has a discrete low pass filter in the following form: 
\begin{equation}
\begin{pmatrix} 
\slfrac{1}{16} & \slfrac{1}{4} & \slfrac{3}{8} & \slfrac{1}{4} & \slfrac{1}{16}\,.
\end{pmatrix}
\end{equation}

\begin{figure}
   \centering
   \includegraphics{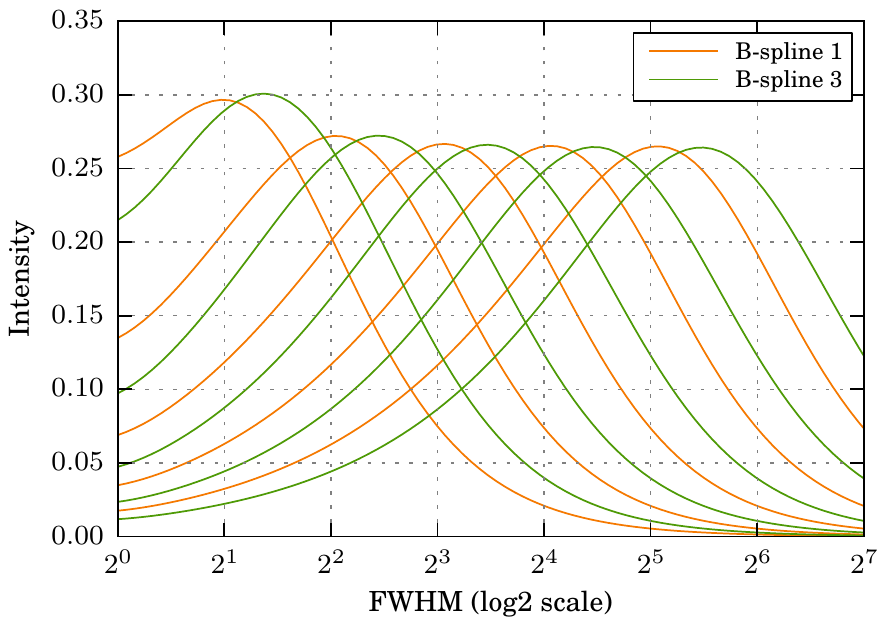}
   \caption{\label{fig:wavelet_response_t_t2} Relation between scales
     of a wavelet transform and spatial scales expressed in units of
     the FWHM of a Gaussian PSF of an image. The individual curves
     represent the relative sensitivity of a given wavelet scale to
     recovering a Gaussian feature with a given FWHM.  The FWHM values
     for which the individual wavelet scales reach their maximum
     sensitivity are marked in the abscissa. The combination of two
     B-spline functions is used in the IWD algorithm to provide the
     enhanced sensitivity over a refined range of spatial scales.}
 \end{figure}

The resulting sensitivity of the IWD to different structural scales is
illustrated in Fig.~\ref{fig:wavelet_response_t_t2} which compares the
responses of different wavelet scales to different spatial scales in the image,
for the scaling functions B-spline and B-spline 3.

\end{appendix}

\end{document}